\documentclass[a4paper,11pt]{article}
\pdfoutput=1
\usepackage[utf8]{inputenc}
\RequirePackage{amsmath}
\RequirePackage{amssymb}
\numberwithin{equation}{section}
\RequirePackage{epsfig}
\RequirePackage{graphicx}
\RequirePackage[numbers,sort&compress]{natbib}
\RequirePackage{color}
\RequirePackage[colorlinks=true
,urlcolor=blue
,anchorcolor=blue
,citecolor=blue
,filecolor=blue
,linkcolor=blue
,menucolor=blue
,linktocpage=true
,pdfproducer=medialab
,pdfa=true
]{hyperref}
\usepackage{authblk}
\usepackage{hyperref}
\usepackage{mathtools}
\usepackage{listings}
\usepackage{mathrsfs}
\usepackage[labelsep=endash,labelfont=bf]{caption}

\usepackage{geometry}
\renewcommand{\d}{\mathrm{d}}

\usepackage{tikz}
\usetikzlibrary{positioning}
\usetikzlibrary{arrows}
\tikzstyle{arrow} = [thin,<->,>=angle 60]
\tikzset{
roundnode/.style={circle, draw, very thick, minimum size=10mm},
squarednode/.style={rectangle, draw, very thick, minimum size=10mm}
}
\tikzset{every loop/.style={}}

\begin{document}

\begin{flushright}
\footnotesize
\texttt{DESY 17-163}
\vspace{2.6cm}
\end{flushright}

%%%%%%%%%%%%%%%%%
\centerline{\LARGE \bf Towards Deconstruction of the}
\medskip

\centerline{\LARGE \bf Type D $\mathbf{(2,0)}$ Theory}

\vspace{.5cm}

 \centerline{\LARGE \bf }

\vspace{1.5truecm}

\centerline{
 {\large \bf Antoine Bourget $^a$} \footnote{bourgetantoine@uniovi.es},
  {\large \bf Alessandro Pini $^b$} \footnote{alessandro.pini@desy.de}   and
    {\large \bf Diego Rodriguez-Gomez $^a$} \footnote{d.rodriguez.gomez@uniovi.es} }

\vspace{1cm}
\centerline{{\it ${}^a$ Department of Physics, Universidad de Oviedo}} \centerline{{\it C/Federico García Lorca 18, 33007, Oviedo, Spain}}
\medskip
\centerline{{\it ${}^b$ DESY Theory Group}}\centerline{{\it Notkestraße 85, 22607 Hamburg, Germany }}
\vspace{2cm}

\centerline{\bf ABSTRACT}
\medskip
%We propose a four-dimensional supersymmetric theory that deconstructs, in a particular limit, the six-dimensional $(2,0)$ theory of type $D_k$. This 4d theory is defined by a necklace quiver with alternating gauge nodes $\mathrm{O}(2k)$ and $\mathrm{Sp}(k)$. We test this proposal by comparing the 6d half-BPS index to the Higgs branch Hilbert series of the 4d theory. The latter is computed using an embedding of the whole construction in string/M-theory, in which a chain of dualities bring us down to three dimensional QFTs, where mirror symmetry lends a helping hand. In the process, we overcome several technical difficulties, among which Hilbert series calculations for non-complete intersections, and the choice of $\mathrm{O}$ versus $\mathrm{SO}$ gauge groups.

We propose a four-dimensional supersymmetric theory that deconstructs, in a particular limit, the six-dimensional $(2,0)$ theory of type $D_k$. This 4d theory is defined by a necklace quiver with alternating gauge nodes $\mathrm{O}(2k)$ and $\mathrm{Sp}(k)$. We test this proposal by comparing the 6d half-BPS index to the Higgs branch Hilbert series of the 4d theory. In the process, we overcome several technical difficulties, such as Hilbert series calculations for non-complete intersections, and the choice of $\mathrm{O}$ versus $\mathrm{SO}$ gauge groups. Consistently, the result matches the Coulomb branch formula for the mirror theory upon reduction to 3d.

\noindent

\newpage

\title{\boldmath Towards Deconstruction of the Type D $(2,0)$ Theory}
\author{Antoine Bourget, Alessandro Pini and Diego Rodr\'iguez-G\'omez}

\tableofcontents

\section{Introduction}

Quantum Field Theories (QFTs) in higher dimensions ($d>4$) have been the object of intensive studies in the very recent past. In particular, the case of six-dimensional (6d) theories stands alone, specially after the seminal work of \cite{Gaiotto:2009we}. Upon compactification on Riemann surfaces, possibly with punctures, infinite new classes of lower dimensional QFTs have been constructed, and many of their properties -- including dualities -- have been understood.

Despite such an enormous progress, very little is known about 6d theories themselves, in particular owing to the lack of a Lagrangian description. The little we know about them comes either from string/M-theory arguments or from inferring from particular limits where other methods are available. Clearly, it would be highly desirable to provide a definition of 6d theories by themselves in purely field-theoretic terms.

The situation is simpler for maximally supersymmetric (SUSY) theories, which come in an $A-D-E$ classification \cite{Witten:1995zh}. Concentrating on the case of type $A$ and type $D$, it has been argued that they can be defined as the UV fixed point of the maximally SUSY theory (MSYM) in 5d with gauge algebra $A$ and $D$ respectively \cite{Douglas:2010iu,Lambert:2010iw}. To explain how this can be possible at all, one notes that 5d Yang-Mills theories contain instanton particles in their spectrum. In the particular case of the maximally SUSY theory it turns out that there is precisely one bound state at threshold at each instanton level. This structure is exactly that of a Kaluza-Klein (KK) tower where the radius of the growing dimension is essentially given by the square of the 5d Yang-Mills coupling constant $g_5^2$. For instance, this can be nicely seen in the instanton contribution to the index in \cite{Kim:2011mv}. Moreover, relying on this connection, in \cite{Kallen:2012zn} it was argued using localization that indeed the maximally SUSY theory with unitary gauge group $\mathrm{SU}(k)$ can host the expected $k^3$ behavior of the free energy.

Another definition of the 6d type $A$ $(2,0)$ theory is provided by the deconstruction mechanism introduced in \cite{ArkaniHamed:2001ca,ArkaniHamed:2001ie}. Interestingly, deconstruction offers the possibility to define the 6d theory through an \textit{a priori} much better controlled 4d theory. The key idea is that, upon going to the equal vacuum expectation value (VEV) Higgs branch in a circular quiver with bifundamental links, the bifundamentals provide a mass to the gauge fields which, in the large number of nodes limit, becomes identical to a KK tower. Thus, an effective extra dimension opens up, and below the scale set by the emergent lattice spacing, the theory enjoys a higher-dimensional Lorentz symmetry. While this mechanism is expected to be generic, it applies in particular to the case of 4d $\mathcal{N}=2$ $\mathrm{U}(k)$ circular quivers with $N$ nodes. In this case, in the deconstruction limit, the 4d theory becomes essentially the 5d MSYM theory. In turn, the latter, and because of the arguments reviewed above, is expected to contain its own UV completion in the instanton sector and hence flow in the UV to the 6d $(2,0)$ theory. 

A similar conclusion can be obtained by embedding the system in string/M-theory. The 4d quiver theory can be regarded as the world-volume theory on a stack of $k$ D3 branes on top of a $\mathbb{C}^2/\mathbb{Z}_N$ singularity. By going to the Higgs branch, we are putting the D3s away from the singularity. Moreover, in the large $N$ limit, the orbifold locally looks like a very thin cylinder. A better description can be found by T-dualizing to IIA, where we find a stack of $k$ D4 branes in flat space, which lift to a stack of $k$ M5 branes whose world-volume hosts the $(2,0)$ type $A_k$ theory, thus confirming the expectation.

Hence, from both perspectives the 4d necklace quiver is expected to deconstruct the 6d $(2,0)$ type $A$ theory. Note that strictly speaking we have both the M-theory circle as well as the one arising from deconstruction, and therefore the 6d theory is living on $\mathbb{R}^4\times \mathbb{T}^2$. 

While deconstruction is a very appealing framework, since it relies on the \textit{a priori} much better controlled 4d quiver theory, very few quantitative tests have been performed. Very recently, relying on supersymmetric localization, two very refined tests of the deconstruction proposal were performed in \cite{Hayling:2017cva}. First, the so-called half-BPS index of the 6d theory was reproduced from a computation in the 4d theory. While this is a non-trivial test of deconstruction, the states counted by the half-BPS index are too simple, in that no state feeling the $\mathbb{T}^2$ --such as self-dual strings-- is counted. In order to be sensitive to those states, as a further check of deconstruction the $\mathbb{S}^4\times \mathbb{T}^2$ partition function was compared between 4d and 6d in \cite{Hayling:2017cva}, obtaining a spectacular match in the deconstruction limit. Thus, these two tests provide very non-trivial evidence of the deconstruction proposal for the $A$-type 6d theory.  

In view of these developments for the type $A$ 6d theory, it is natural to wonder about the extension of deconstruction to the $D$-type case. In this paper we will take the first steps in this direction, using the half-BPS index as a sanity check for our proposal. As argued in \cite{Hayling:2017cva}, the half-BPS index of the 6d theory is captured by the counting of chiral operators in the Higgs branch of the 4d quiver. Since the Higgs branch does not change upon dimensional reduction, we can compute it in the 3d version of the deconstructing theory. The key observation is that one can now use mirror symmetry and count Coulomb branch operators in the ``magnetic" theory. As we will describe below, it turns out that this computation is much simpler and provides a direct inspiration for a theory mirror to a candidate to deconstruct the $D$-type theory. By construction this theory is designed to reproduce the correct half-BPS index. Moreover, it turns out that this theory can be engineered in string/M-theory on the world-volume of a stack of branes which corresponds to what one would have naively guessed \textit{a priori} to deconstruct $D$-type theories, namely, the same set-up as for type $A$ only with the extra addition of an orientifold plane. Thus, we find consistency among the string/M-theory picture and the localization test of the matching of the half-BPS index.

The rest of the paper is organized as follows. In section \ref{sectionDeconstruction} we briefly review some of the salient aspects of the deconstruction proposal of \cite{ArkaniHamed:2001ca,ArkaniHamed:2001ie}, with special emphasis on the string/M-theory embedding. Following \cite{Hayling:2017cva}, we introduce the half-BPS index as the simplest diagnostics tool of deconstruction, and review its computation in the 4d deconstructing theory through the so-called Higgs branch Hilbert series. In section \ref{Mirror} we turn to the computation of the half-BPS index through mirror symmetry upon reduction to 3d. Very generically we can argue that the 3d theory mirror to a candidate to deconstruct the 6d theory should be a $\mathrm{O}(2k)$ theory with an adjoint hypermultiplet and $2N$ vector half-hypermultiplets. Regarding this theory as a particular element in the class of theories based on a classical gauge group $G$ of B, C and D type, we investigate the mirror for each of these. One subtle point is whether one should consider full orthogonal $\mathrm{O}(n)$ or special orthogonal $\mathrm{SO}(n)$ groups, a question we solve using precise Hilbert series techniques. As a by-product, we develop a strategy to compute the Higgs branch Hilbert series for non-complete intersections efficiently by using letter-counting in an auxiliary theory with an extended matter sector. In particular, we find that the candidate to deconstruct the $D$-type 6d theory is a circular quiver with $[\mathrm{O}(2k)\times \mathrm{Sp}(k)]^N$ gauge groups connected by half-hypermultiplets. In section \ref{secDeconstructionD} we argue that this theory is indeed the one which one would most naively have guessed, in view of a string/M-theory construction, as candidate to deconstruct the $D$-type 6d theory and comment on some open questions raised by this conjecture. Finally, we end with some conclusions in section \ref{conclusions}. We compile in the appendices \ref{A1} and \ref{AppendixProof} several technical results which play a role in the computation of the mirror pairs of section \ref{Mirror}.

\section{Deconstruction of the type A theory}
\label{sectionDeconstruction}

In \cite{ArkaniHamed:2001ie} it was proposed that the 4d $\mathcal{N}=2$ circular quiver theory with $N$ $\mathrm{U}(k)$ nodes joined by bifundamental hypermultiplets -- in the following we will denote this theory $\mathscr{B}^N_{\mathrm{U}(k)}$, see Figure \ref{figureTheoriesAB} -- deconstructs, upon going to the (equal VEV $v$) Higgs branch and upon taking large $N$, the type $A_k$ 6d $(2,0)$ theory on $\mathbb{R}^4\times \mathbb{T}^2$.

Let us briefly review how this comes about (we refer to \cite{ArkaniHamed:2001ca,ArkaniHamed:2001ie} for a detailed account). The $\mathbb{T}^2$ spans the $x^5$ and $x^6$ directions. The $x^5$ direction of the torus is generated by the deconstruction mechanism, and its radius is given by 

\begin{equation}
2\pi R_5=\frac{N}{G\,v}\, , 
\end{equation}
being $v$ the Higgs VEV and $G$ the 4d gauge coupling -- equal to all nodes. Moreover, it comes with a lattice spacing $a = (Gv)^{-1} $, so that at distances large compared to $a$, the theory behaves as an approximately Lorentz invariant discretized 5d gauge theory with  gauge coupling 

\begin{equation}
    g_5^2=\frac{G}{v}\, .
\end{equation}
Then, assuming that the theory does not generate an IR scale, and that there is no phase transition as $G \rightarrow \infty$, we can consider the large $N,G,v$ limit where $g_5$ and $R_5$ are kept fixed while $a \rightarrow 0$. In this limit, the 4d quiver becomes equivalent at all scales to the maximally SUSY Yang-Mills theory in 5d (see \cite{Lambert:2012qy} for a detailed description). 

But there remains another scale in the theory set by the 5d YM coupling $g_5^{-2}$. For energies $\ll g_5^{-2}$, the resultant theory reproduces 5d MSYM on a circle of radius $R_5$ and bare gauge coupling $g_5$. However, the 5d MSYM can be interpreted as the low-energy description for 6d $\mathcal{N}=(2,0)$ theory on a circle of radius $R_6$
\begin{equation}
    2\pi R_6=g_5^2\, .
\end{equation}
The KK modes of the reduction correspond to the instantons of the 5d theory, which hence contains its UV completion. At strong coupling, the KK modes become light and the theory becomes effectively the 6d $(2,0)$ type A theory.

The discussion of deconstruction can be embedded in string/M-theory. The deconstructing theory $\mathscr{B}^N_{\mathrm{U}(k)}$ can be regarded as the world-volume theory on a stack of $k$ D3 branes transverse to $\mathbb{C}\times\mathbb{C}^2/\mathbb{Z}_N$. Going to the Higgs branch amounts to locating the branes somewhere in the $\mathbb{C}^2/\mathbb{Z}_N$ far away from its tip. In turn, the large $N$ limit makes the orbifold to look locally like a very thin cylinder. A better description is then found by T-dualizing, thus finding $k$ D4 branes in flat space -- whose world-volume theory is the emerging 5d MSYM. Then, at strong coupling, the D4 branes uplift to $k$ M5 branes, which host the 6d $(2,0)$ theory.

Reverse-engineering the set-up, we can start with an M-theory configuration with $k$ M5 branes along $(x^1,x^2,x^3,x^6,x^{10})$ and $N$ M5 branes along $(x^1,x^2,x^3,x^4,x^5)$. Reducing on $x^{10}$ gives a type IIA configuration with $k$ D4 branes along $(x^1,x^2,x^3,x^6)$ and $N$ NS5 branes along $(x^1,x^2,x^3,x^4,x^5)$, from which the $\mathscr{B}^N_{\mathrm{U}(k)}$ is easily read-off. Further $T$-duality along $x^6$ gives the original picture in terms of $k$ D3 branes probing a $\mathbb{C}^2/\mathbb{Z}_N$ singularity along $(x^6,x^7,x^8,x^9)$. Then, deconstruction proceeds just as described above.

%In this language, the deconstruction procedure amounts to placing the D3 branes far away from the tip of the cone, which becomes very thin in the deconstruction limit. A better description is then the IIA set-up, where the D4 branes wrapping the $x^6$ circle join and detach from the NS5's in the transverse directions. We thus have $k$ D4 branes in flat space, whose UV description is in terms of $k$ M5 branes in flat space, far away from the other $N$ M5 which generate the orbifold in the IIB description.

As we will discuss in the next subsection, a first probe of deconstruction is the computation of the half-BPS index of the 6d theory through the deconstructing 4d theory $\mathscr{B}^N_{\mathrm{U}(k)}$. In the latter, the relevant objects to count are chiral gauge invariant operators in the Higgs branch. The Higgs branch of theories with eight supercharges is independent of the dimension of the theory (4d or 3d) \cite{Argyres:1996eh}, so we can as well compute it for the theory obtained from the IIA set-up by reducing along $x^3$. This is implemented by T-duality, obtaining a IIB configuration with $k$ D3 branes along $(x^1,x^2,x^6)$ and $N$ NS5 branes along $(x^1,x^2,x^3,x^4,x^5)$. It is clear that the quiver is just the same $\mathscr{B}^N_{\mathrm{U}(k)}$, and that the Higgs branch is identical to the 4d case.

The further reduction to 3d might look trivial, but on the contrary it allows us for a new possibility. Namely, we can now use mirror symmetry and claim that the Hilbert series of the Coulomb branch of the mirror also reproduces the 6d 1/2 BPS index. Such mirror is a $\mathrm{U}(k)$ gauge theory with N flavors. To see this we can start from the M-theory configuration and reduce along dimension 3 --note that the 10d coordinates are then $(x^0,x^1,x^2,x^4,x^5,x^6,x^7,x^8,x^9,x^{10})$--, and then perform $T_{10}$-duality, to obtain $k$ D3 branes along $(x^1,x^2,x^6)$ and $N$ D5 branes along $(x^1,x^2,x^3,x^4,x^5)$, which, as expected, is nothing but the S-dual of the IIB 3d configuration. One further T-duality to IIA along $x^6$ gives $k$ D2 branes along $(x^1,x^2)$ and $N$ D6 branes along $(x^1,x^2,x^4,x^5,x^6,x^{10})$, from which we can easily read off the mirror theory in terms of a $\mathrm{U}(k)$ theory with N hypermultiplets.

\begin{figure}[t]
    \centering
    \def\x{2}
\begin{tikzpicture}[scale=0.75,every node/.style={scale=0.75}]
\node[draw,text width=3cm,align=center] (1) at (4*\x,4*\x) {M-Theory \\ $k$ M5$_{0,1,2,3,6,10}$ \\ $N$ M5$_{0,1,2,3,4,5}$};
\node[draw,text width=3cm,align=center] (2) at (2.5*\x,2*\x) {IIA \\ $k$ D4$_{0,1,2,3,6}$ \\ $N$ NS5$_{0,1,2,3,4,5}$};
\node[draw,text width=3cm,align=center] (3) at (5.5*\x,2*\x) {IIA \\ $k$ D4$_{0,1,2,6,10}$ \\ $N$ D4$_{0,1,2,4,5}$};
\node[draw,text width=3cm,align=center] (4) at (0*\x,2*\x) {IIB \\ $k$ D3$_{0,1,2,3}$ \\ $\mathbb{C}^2/\mathbb{Z}_N$ on $6,7,8,9$};
\node[draw,text width=3cm,align=center] (6) at (2.5*\x,0*\x) {IIB \\ $k$ D3$_{0,1,2,6}$ \\ $N$ NS5$_{0,1,2,3,4,5}$};
\node[draw,text width=3cm,align=center] (7) at (5.5*\x,0*\x) {IIB \\ $k$ D3$_{0,1,2,6}$ \\ $N$ D5$_{0,1,2,4,5,10}$};
\node[draw,text width=3cm,align=center] (8) at (0*\x,0*\x) {IIA \\ $k$ D2$_{0,1,2}$ \\ $\mathbb{C}^2/\mathbb{Z}_N$ on $6,7,8,9$};
\node[draw,text width=3cm,align=center] (9) at (8*\x,0*\x) {IIA \\ $k$ D2$_{0,1,2}$ \\ $N$ D6$_{0,1,2,4,5,6,10}$};
\draw [arrow] (1) -- (2) node[midway,left] {$\odot_{10}$};
\draw [arrow] (1) -- (3) node[midway,right] {$\odot_{3}$};
\draw [arrow] (2) -- (4) node[midway,above] {$T_6$};
\draw [arrow] (6) -- (8) node[midway,above] {$T_6$};
\draw [arrow] (9) -- (7) node[midway,above] {$T_6$};
\draw [arrow] (2) -- (6) node[midway,left] {$T_3$};
\draw [arrow] (4) -- (8) node[midway,left] {$T_3$};
\draw [arrow] (3) -- (7) node[midway,left] {$T_{10}$};
\draw [arrow] (6) -- (7) node[midway,below] {$S$/Mirror};
\end{tikzpicture}
    \caption{Summary of the dualities. The top node is the M-theory node, the second line contains the 4d theories (plus the D4/D4) and the third line the 3d theories. The symbol $\odot$ means that we shrink the associated circle to zero size. }
    \label{summaryDualities}
\end{figure}
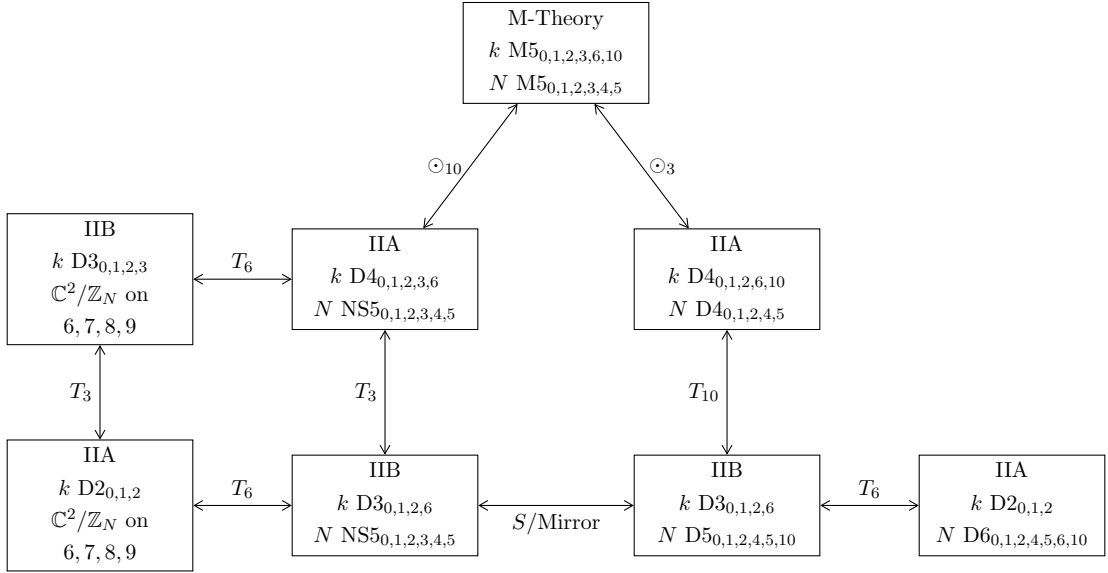

For concreteness, we summarize the chains of dualities in Figure \ref{summaryDualities}.

In the following we will be mostly concentrated on the 3d avatar. As described above, the ``electric theory" is generically a necklace quiver. In the type $A$ case described above such necklace is the familiar quiver theory with N $\mathrm{U}(k)$ nodes (more generically we could write a generic group $G$) connected by bifundamentals which we dubbed $\mathscr{B}^N_{G}$. However, in the following a necklace with $N$ copies of $G_1\times G_2$ joined for some groups $G_i$ will be relevant. $G_i$ will be either a symplectic or an (special) orthogonal group, and the basic structure $G_1\times G_2$ will alternate those. Therefore, the links are in this case half-hypermultiplets. We will refer to such theory $\mathscr{B}^N_{G_1,G_2}$. In addition we will need the ``magnetic theories" to those. These are generically a 3d theory with gauge group $G$, one adjoint hypermultiplet and $N$ matter fields. Note that in the case when $G$ is symplectic/(special) orthogonal, the matter fields will be half-hypermultiplets.

\begin{figure}[t]
\centering
\begin{tabular}{|c|c|}
\hline
    Name & Quiver  \\ \hline 
  $\mathscr{A}^N_{G}$  &\begin{tikzpicture}[node distance=2cm]
\node[roundnode] (1) {$G$};
\node[squarednode] (2) [right of=1] {$N$};
\draw[-] (1) -- (2);
\draw (1) edge [loop left] node {Adj} (1);
\end{tikzpicture} \\ \hline
    $\mathscr{B}^N_{G}$  & 
         \def\x{2}
\begin{tikzpicture}
\node (0) at (0*\x,0*\x) {$N$ nodes};
\node[roundnode] (1) at (1*\x,0*\x) {$G$};
\node[roundnode] (2) at (.707*\x,.707*\x) {$G$};
\node[roundnode] (3) at (0*\x,1*\x) {$G$};
\node[roundnode] (4) at (-.707*\x,.707*\x) {$G$};
\node[roundnode] (5) at (-1*\x,0*\x) {$G$};
\node[roundnode] (6) at (-.707*\x,-.707*\x) {$G$};
\node[roundnode] (7) at (0*\x,-1*\x) {$G$};
\node[roundnode] (8) at (.707*\x,-.707*\x) {$G$};
\draw[-] (1) -- (2);
\draw[-] (2) -- (3);
\draw[-] (4) -- (3);
\draw[-] (4) -- (5);
\draw[-] (6) -- (5);
\draw[-] (6) -- (7);
\draw[-] (8) -- (7);
\draw[-] (8) -- (1);
\end{tikzpicture}
 \\ \hline 
   $\mathscr{B}^N_{G_1,G_2}$  & 
         \def\x{2}
\begin{tikzpicture}
\node (0) at (0*\x,0*\x) {$2N$ nodes};
\node[roundnode] (1) at (1*\x,0*\x) {$G_1$};
\node[roundnode] (2) at (.707*\x,.707*\x) {$G_2$};
\node[roundnode] (3) at (0*\x,1*\x) {$G_1$};
\node[roundnode] (4) at (-.707*\x,.707*\x) {$G_2$};
\node[roundnode] (5) at (-1*\x,0*\x) {$G_1$};
\node[roundnode] (6) at (-.707*\x,-.707*\x) {$G_2$};
\node[roundnode] (7) at (0*\x,-1*\x) {$G_1$};
\node[roundnode] (8) at (.707*\x,-.707*\x) {$G_2$};
\draw[-] (1) -- (2);
\draw[-] (2) -- (3);
\draw[-] (4) -- (3);
\draw[-] (4) -- (5);
\draw[-] (6) -- (5);
\draw[-] (6) -- (7);
\draw[-] (8) -- (7);
\draw[-] (8) -- (1);
\end{tikzpicture}
 \\ \hline 
\end{tabular}
    \caption{Definition of theories used in the article.}
    \label{figureTheoriesAB}
\end{figure}

\subsection{The Half BPS Index}\label{halfBPS}

A natural way to quantitatively test the deconstruction proposal is to identify a subsector of operators with nice properties affording a counting in both the 6d and 4d deconstructing theory. Then, the matching of such partition functions can then be interpreted as a consistency check for the deconstruction proposal \cite{Hayling:2017cva}.

The maximal compact subalgebra of the 6d superconformal algebra $\mathfrak{osp}(8^*|4)$ is $\mathfrak{so}(6)\oplus \mathfrak{so}(2)\oplus\mathfrak{so}(5)_R$, whose Cartan operators will be denoted respectively as $(h_1,\,h_2,\,h_3)$ for the Lorentz symmetry and $(J_1,\,J_2)$ for the R-symmetry --we follow the conventions in \cite{Buican:2016hpb}. In addition, the fermionic generators are $\mathcal{Q}_{\mathbf{A}a}$ and $\mathcal{S}_{\mathbf{A}\dot{a}}$, where $a,\,\dot{a}$ are the Lorentz spinor indices and $\mathbf{A}$ the R-symmetry (spinor) index. 

In turn, the 4d superconformal algebra $\mathfrak{su}(2,2|2)$ has as maximal compact subalgebra $\mathfrak{so}(4)\oplus \mathfrak{so}(2)\oplus\mathfrak{su}(2)_R\oplus \mathfrak{u}(1)_r$, whose Cartan subalgebras will be denoted as $(m_1,\,m_2)$ for the Lorentz part and $(R,r)$ for the R-symmetry part. In addition, the supercharges will be denoted as $Q^I_{\alpha},\,\widetilde{Q}_{I\dot{\alpha}}$, where now $I$ is the $\mathfrak{su}(2)_R$ index and $\alpha,\,\dot{\alpha}$ the Lorentz spinor index.

Relying on \cite{Beem:2014kka}, the 4d superconformal algebra can be embedded into the 6d one in such a way that 
\begin{equation}
\label{Qmap}
    \mathcal{Q}_{\mathbf{1},1},\, \mathcal{Q}_{\mathbf{1},2} \leftrightarrow  Q^1_{\alpha}\, ; \qquad 
    \mathcal{Q}_{\mathbf{2},3},\, \mathcal{Q}_{\mathbf{2},4}\leftrightarrow  \widetilde{Q}_{2\dot{\alpha}}\, .
\end{equation}
As shown in \textit{e.g.} \cite{Beem:2014kka,Buican:2016hpb}, the primaries annihilated by $\mathcal{Q}_{\mathbf{1}a},\,\mathcal{Q}_{\mathbf{2}a}$ in addition to the $\mathcal{S}_{\mathbf{A}a}$ give rise to the $\mathcal{D}[0,0,0;J_1-J_2,0]$ multiplets. Their primaries are counted by the so-called half-BPS index. In turn, owing to $\eqref{Qmap}$, these translate to states in 4d annihilated by $Q^1_{\alpha},\,\widetilde{Q}_{2\dot{\alpha}}$. Such shortening condition, corresponding to the so-called $\hat{\mathcal{B}}_R$ multiplets in the notation of \cite{Dolan:2002zh}, defines Higgs branch operators. The counting of these can be done through the so-called Higgs branch Hilbert series \cite{Benvenuti:2010pq}. Thus, all in all, we conclude that the 6d half-BPS index should be captured by the 4d Higgs branch Hilbert series. Note that such expectation is based solely on algebraic reasons, and hence is expected to hold not only for $A$-type theories as in \cite{Hayling:2017cva}, but in general.

The spectrum of half-BPS operators, at least for type $A$ theories, has been thoroughly studied in the literature (see \textit{e.g.} \cite{Aharony:1997th,Aharony:1997an,Bhattacharyya:2007sa}, or for a more recent account \cite{Beem:2014kka}. The localization derivation appears in \cite{Kim:2012tr}). In particular, it turns out that the half-BPS operators are in one-to-one correspondence with the Casimir operators of the corresponding group. For the $A$ type with group $U(k)$, this leads to the well-known result
\begin{equation}
\label{IA}
    \mathcal{I}_{\frac{1}{2}{\rm BPS}}^{A_k (2,0)} := \mathcal{I}_{\frac{1}{2}{\rm BPS}}^{U(k)}={\rm PE}\left[\sum_{i=1}^k\,u^i\right]\,.
\end{equation}
Note that dividing by the $k=1$ case gives the counting of Casimirs of the $SU(k)$ case, which can be interpreted as removing the free tensor multiplet associated to the center of mass. This interpretation does not extend to the $D$-type case. Hence, it is natural to assume the $D$-type theory to have gauge group $\mathrm{O}(2k)$, so that the half-BPS index should be\footnote{Recall for comparison that \begin{equation}
    \mathcal{I}_{\frac{1}{2}{\rm BPS}}^{\mathrm{SO}(2k)}={\rm PE}\left[u^k + \sum_{i=1}^{k-1}\,u^{2i}\right]\,.
\end{equation}}
\begin{equation}
\label{ID}
    \mathcal{I}_{\frac{1}{2}{\rm BPS}}^{D_k  (2,0)} := \mathcal{I}_{\frac{1}{2}{\rm BPS}}^{\mathrm{O}(2k)}={\rm PE}\left[\sum_{i=1}^k\,u^{2i}\right]\,.
\end{equation}
This expression, arising from assuming $\mathrm{O}(2k)$ gauge group, has the very interesting property that $ \mathcal{I}_{\frac{1}{2}{\rm BPS}}^{D_k  (2,0)}$ can be interpreted as the $k$-fold symmetrized product of the $k=1$ case. More explicitly
\begin{equation}
\label{abelianization}
    {\rm PE}\left[\nu \frac{1}{1-u^2}\right] = \sum_{k=0} \nu^k\,\mathcal{I}_{\frac{1}{2}{\rm BPS}}^{\mathrm{O}(2k)}\, .
\end{equation}
Thus, just like the type $A$ case \cite{Hayling:2017cva}, the $D$-type 6d theory index with gauge group $\mathrm{O}(2k)$ ``abelianizes".

Note that, as discussed in \cite{Hayling:2017cva}, the half-BPS index is a rough observable in that it is only sensible to perturbative local operators and not to states which feel the $\mathbb{T}^2$ -- such as wrapped self-dual strings. However, in the following we will use the half-BPS index as a guide towards the deconstruction of the 6d theory, as it is the simplest non-trivial quantitative check of any such proposal.

\subsection{Computation of the Higgs branch Hilbert series}
\label{SectionComputationU}

In this section, we will explain how to compute the Higgs branch Hilbert series for the quiver theory $\mathscr{B}^N_{\mathrm{U}(k)}$. The standard computation of the Higgs branch Hilbert series proceeds just as in \cite{Benvenuti:2010pq}. However, special care has to be taken when the Higgs branch is not a complete intersection, which is typically the case for the quivers of the form of $\mathscr{B}^N_{G}$ (see \cite{Dey:2013fea,Hayling:2017cva} for the type $A$). In the case of the type $A$ one can use brute force methods and compute, with the help of computer packages, the relevant Hilbert series. However, it turns out that it is possible to devise an algorithm allowing to compute analytically the desired result. To be precise, we will proceed in three steps: 

\begin{enumerate}
    \item We define an auxiliary theory $\tilde{\mathscr{B}}^N_{\mathrm{U}(k)}$ by the quiver of Figure \ref{FigBtildeTheoryU}, and show that it has the same Higgs branch Hilbert series as the original theory; 
    \item We prove that the F-terms variety of $\tilde{\mathscr{B}}^N_{\mathrm{U}(k)}$ is a complete intersection; 
    \item We use step 2 to compute the Higgs branch Hilbert series of $\tilde{\mathscr{B}}^N_{\mathrm{U}(k)}$, and therefore $\mathscr{B}^N_{\mathrm{U}(k)}$. 
\end{enumerate}
Let us now go into more detail for each step. 
\begin{enumerate}
    \item The Higgs branch Hilbert series counts chiral gauge-invariant operators on the Higgs branch, graded by their conformal dimension. For theory $\mathscr{B}^N_{\mathrm{U}(k)}$, these operators are built from the hypermultiplets. Let us denote by $\Phi_i$ the vector multiplet in the node $i$ of the quiver, for $i \in \mathbb{Z}_N$. Then the superpotential reads 
    \begin{equation}
        W = \mathrm{Tr} \left( \sum\limits_{i \in \mathbb{Z}_N} X_i \Phi_{i+1} Y_i - Y_i \Phi_{i} X_i \right) = 
            \mathrm{Tr} \left( \sum\limits_{i \in \mathbb{Z}_N} X_{i-1} \Phi_{i} Y_{i-1} - Y_i \Phi_{i} X_i \right) \, . 
    \end{equation}
    Here the $X_i$ and $Y_i$ are the hypermultiplets that transform in the representations $\mathrm{fund}_i \times \overline{\mathrm{fund}}_{i+1}$ and $\mathrm{fund}_{i+1} \times \overline{\mathrm{fund}}_{i}$ respectively. 
    The relevant F-terms are 
    \begin{equation}
        Y_{i-1} X_{i-1} - X_i Y_i = 0 \, , \qquad i \in \mathbb{Z}_N \, . 
    \end{equation}
    Now in the $\tilde{\mathscr{B}}^N_{\mathrm{U}(k)}$ theory, we have an additional pair of hypermultiplets $Z$ and $Z'$ that transform in the fundamental and antifundamental of the node number $1$. The superpotential then has an additional contribution $Z' \Phi_{1} Z $, and the relevant F-term for $i=1$ is modified:  
    \begin{eqnarray}
    \label{FtermsBtildeU}
    Y_{i-1} X_{i-1} - X_i Y_i &=& 0 \, , \qquad i \in \mathbb{Z}_N - \{1\} \, , \\ \nonumber
        Y_{0} X_{0} - X_1 Y_1 + ZZ' &=& 0 \, . 
    \end{eqnarray}
    Using this, we can construct a bijection between gauge invariant operators of the two theories. We can focus on the single trace operators. There is an obvious identity map 
    \begin{equation}
        \mathrm{GIO}^{\mathrm{st}} \left[ \mathscr{B}^N_{\mathrm{U}(k)} \right] \rightarrow \mathrm{GIO}^{\mathrm{st}}  \left[ \tilde{\mathscr{B}}^N_{\mathrm{U}(k)} \right] \, ,
    \end{equation}
    where $\mathrm{GIO}^{\mathrm{st}}$ stands for gauge-invariant operators that can be written as a single trace. This map is surjective, since any operator in $\mathrm{GIO}^{\mathrm{st}}  \left[ \tilde{\mathscr{B}}^N_{\mathrm{U}(k)}\right]$ that involves the fields $Z$ and $Z'$ need to be of the form $Z' O Z$ where $O$ is a product of $X_i$ and $Y_i$ fields. But then 
    \begin{equation}
    \label{ZZelimination}
        Z' O Z = \mathrm{Tr} \left( OZZ' \right) =  \mathrm{Tr} \left( OX_1 Y_1 \right) -   \mathrm{Tr} \left( OY_{0} X_{0}\right) \, . 
    \end{equation}
    Moreover, the map is an injection, because the process (\ref{ZZelimination}) of $Z$ and $Z'$ elimination is unique. 
    Therefore, we have proved that the single-trace operators of the two theories can be put in bijection, and the same holds for any operators. We deduce that the Hilbert series of the two theories are equal. 
    \begin{figure}[t]
    \centering
         \def\x{2}
\begin{tikzpicture}
\node[squarednode] (-1) at (-2*\x,0*\x) {$\mathrm{U}(1)$};
\node (0) at (0*\x,0*\x) {$N$ nodes};
\node[roundnode] (1) at (1*\x,0*\x) {$\mathrm{U}(k)$};
\node[roundnode] (3) at (0*\x,1*\x) {$\mathrm{U}(k)$};
\node[roundnode] (5) at (-1*\x,0*\x) {$\mathrm{U}(k)$};
\node[roundnode] (7) at (0*\x,-1*\x) {$\mathrm{U}(k)$};
\draw[-] (-1) -- (5);
\draw[-] (1) -- (3);
\draw[-] (3) -- (5);
\draw[-] (5) -- (7);
\draw[-] (7) -- (1);
\end{tikzpicture}
    \caption{Quiver defining the $\tilde{\mathscr{B}}^N_{\mathrm{U}(k)}$ theory. }
    \label{FigBtildeTheoryU}
\end{figure}
    \item Having argued for the equality of the Higgs branches of the original and the extended theory, we can now concentrate on the extended theory. As usual, the computation of the Higgs branch Hilbert series proceeds by first enumerating all possible monomials modulo F-terms and then projecting onto gauge invariants. Let us focus on the first problem. To that matter, we have to analyze the F-terms (\ref{FtermsBtildeU}) of the extended theory forgetting for the time being the gauge integration -- which is, of course, to be performed as the last step. It turns out out that they define a complete intersection. To see that, we can solve the equations formally. First we start with 
      \begin{equation}
     Y_i = \left(\prod\limits_{j=2}^{i} X_j\right)^{-1} Y_{1} \left(\prod\limits_{j=1}^{i-1} X_j\right) \, , \qquad i \in \mathbb{Z}_N - \{1\} \, .
    \end{equation}
    In particular, we have 
    \begin{equation}
     Y_{0} = Y_{N} = \left(\prod\limits_{j=2}^{N} X_j\right)^{-1} Y_{1} \left(\prod\limits_{j=1}^{N-1} X_j\right)  \, .
    \end{equation}
    The last equation becomes, denoting $\mathbf{X} = \prod\limits_{j=1}^{N} X_j$ and $\mathbf{Z}=\mathbf{X} X_1^{-1} ZZ' $,  
    \begin{equation}
         Y_1 \mathbf{X}- \mathbf{X} Y_1 + \mathbf{Z}= 0 \, . 
    \end{equation}
    This equation defines a complete intersection by the result of section \ref{secExCI}, and this completes the proof. 
    Note that for the theory $\mathscr{B}^N_{\mathrm{U}(k)}$ the last equation would read $ Y_1 \mathbf{X}- \mathbf{X} Y_1= 0 $, and this equation does not define a complete intersection, by the result of section \ref{secExNCI}. 
    \item Now we make the projection onto gauge invariants. For $k=1$, we have to evaluate the integral 
    \begin{equation}
       \mathcal{H}[\tilde{\mathscr{B}}^N_{\mathrm{U}(1)}] = \prod\limits_{i \in \mathbb{Z}_N}\oint_{\mid z_i \mid = 1}  \frac{\mathrm{d} z_i}{2 \pi i z_i} \mathrm{PE} \left[-N t^2 + t \left(\frac{z_1}{u} + \frac{u}{z_1}\right) + t \sum\limits_{i\in \mathbb{Z}_N} \left(\frac{z_i}{z_{i+1}} + \frac{z_{i+1}}{z_{i}} \right) \right] \, , 
    \end{equation}
    where the $z_i$ are the fugacities for the different  $\mathrm{U}(1)$ gauge nodes and $u$ is the fugacity for the flavour $\mathrm{U}(1)$ node. One can show (see Appendix \ref{AppendixProof} for the details of the calculation) that
    \begin{equation}
    \label{HS1}
        \mathcal{H}[\tilde{\mathscr{B}}^N_{\mathrm{U}(1)}] = \mathrm{PE}\left[t^2 + 2 t^{N} - t^{2N}\right] \, . 
    \end{equation}
Then for higher values of $k$, we use the ``abelianization" trick, namely 
\begin{equation}
       \mathrm{PE}\left[ \nu \mathcal{H}[\mathscr{B}^N_{\mathrm{U}(1)}] \right] = 1+ \sum\limits_{k=1}^{\infty} \nu^k \mathcal{H}[\mathscr{B}^N_{\mathrm{U}(k)}] \, . 
    \end{equation}
\end{enumerate}

We can offer a physical argument in support of the procedure which we have just described \cite{Dey:2013fea,Hayling:2017cva}. Upon adding $F_i$ flavors to node $i$ to the $\mathscr{B}^N_{\mathrm{U}(k)}$, the Higgs branch of the resulting theory constructs the moduli space of $k$ instantons of the unitary group of rank $\sum F_i$ on $\mathbb{C}^2/\mathbb{Z}_N$ (see \cite{Dey:2013fea} and references therein for details). The case we are really interested in is $F_i=0$, which would correspond $k$ ``rank zero" instantons. Each of them can be thought as a point particle, with no other degree of freedom than those purely geometric. This heuristically explains why the Higgs branch is the $k$-fold symmetrized product of $\mathbb{C}^2/\mathbb{Z}_N$. In turn, $\tilde{\mathscr{B}}^N_{\mathrm{U}(k)}$ corresponds to, say, $F_1=1$ while all others vanishing. This case corresponds to $\mathrm{U}(1)$ instantons, which also behave as a point particles as they cannot have internal degrees of freedom. Hence, both the $\mathscr{B}^N_{\mathrm{U}(k)}$ and the $\tilde{\mathscr{B}}^N_{\mathrm{U}(k)}$, have the same Higgs branch equal to the $k$-fold symmetrized product of $\mathbb{C}^2/\mathbb{Z}_N$.

\section{Mirror Symmetry and Hilbert Series}
\label{Mirror}

In section \ref{sectionDeconstruction}, we have reviewed how the theory $\mathscr{B}^{N}_{\mathrm{U}(k)}$ deconstructs the six-dimensional $\mathcal{N}=(2,0)$ theory of type $A_k$. One exact check of this statement is the comparison of the Higgs branch Hilbert series of the four-dimensional theory with the six-dimensional half-BPS index in the large $N$ limit. As the Higgs branch is not modified upon dimensional reduction, we may as well consider, for this matter, the 3d version of the $\mathscr{B}^{N}_{\mathrm{U}(k)}$ theory. In turn, using mirror symmetry, we can likewise compute it through the Coulomb branch of the magnetic theory. Therefore let's consider the $\mathscr{A}^N_{\mathrm{U}(k)}$ theory introduced in section \ref{sectionDeconstruction}. 

Due to 3d mirror symmetry, the Coulomb branch Hilbert series of $\mathscr{A}^N_{\mathrm{U}(k)}$ must be equal to the Higgs branch Hilbert series of $\mathscr{B}^N_{\mathrm{U}(k)}$. In turn, as explained in \cite{Cremonesi:2013lqa}, the Coulomb branch Hilbert series of $\mathscr{A}^N_{\mathrm{U}(k)}$ reads
\begin{equation}
\label{eq:hsuk}
    \mathcal{C}\left[\mathscr{A}^N_{\mathrm{U}(k)}\right] = \sum_{m_1 \geq m_2 \geq ... \geq m_N > - \infty} t^{2\Delta(\vec{m})}P_{\mathrm{U}(N)}(t^2;\vec{m}) \, ,
\end{equation}
where $\vec{m}=(m_1,...,m_N)$ are the magnetic fluxes, while $\Delta(\vec{m})$ is the conformal dimension of the monopole operator of that flux, that is a function of the matter content of the quiver gauge theory. Finally the classical factor $P_{\mathrm{U}(N)}$ can be expressed as follow. As explained in appendix A of \cite{Cremonesi:2013lqa} we can associate to $\vec{m}$ a partition $\lambda(\vec{m})$ of $N$, such that $\sum_{i}\lambda_{i}(\vec{m}) = N$ and $\lambda_{i}(\vec{m}) \geq \lambda_{i+1}(\vec{m}) $. This partition tells us how many of the fluxes are equal. Using this partition the factor $P_{\mathrm{U}(N)}$ reads \cite{Cremonesi:2013lqa}
\begin{equation}
    P_{\mathrm{U}(N)}(t;m) = \prod_{i=1}^{N}Z^{U}_{\lambda_{i}(\vec{m})} \, ,
\end{equation}
where
\begin{equation}
    Z_{k}^{U} = \prod_{i=1}^{k}\frac{1}{1-t^i} \ \ \textrm{for} \ \ k \geq 1 \ \ \ \ \textrm{and} \ \ \ \ Z^{U}_{0} = 1 \, .
\end{equation}
From a physical point of view the classical factor $P_{\mathrm{U}(N)}$ is counting the Casimir invariants of the residual gauge group.

Let's now consider the large $N$ limit. We see that the only contribution to the Hilbert series (\ref{eq:hsuk}) comes from the configuration $\vec{m}=(0,...,0)$. Therefore we obtain
\begin{equation}
\lim_{N \to + \infty}\mathcal{C}\left[\mathscr{A}^N_{\mathrm{U}(k)}\right] = P_{\mathrm{U}(N)}(t^2;0,...,0) =  \textrm{PE}\left[\sum_{j=1}^{k}t^{2j}\right] \, , 
\end{equation}
which is the half-BPS index of (2,0) 6d theory of type  $A_k$ in \eqref{IA}, using $u=t^2$. 

The previous computation of the Coulomb branch Hilbert series can be easily extended also for other kinds of gauge group. In general the Coulomb branch Hilbert series for a 3d $\mathcal{N}=4$ of the type $\mathscr{A}^N_{G}$ reads \cite{Cremonesi:2013lqa}
\begin{equation}
    \label{eq:hsgeneral}
    \mathcal{C}\left[\mathscr{A}^N_{G}\right] = \sum_{m \ \in \ \Gamma^{\star}_{\hat{G}}/W_{\hat{G}}} t^{2\Delta(\vec{m})}P_{G}(t^2;\vec{m}) \, ,
\end{equation}
where $W_{\hat{G}}$ denotes the Weyl group of the dual GNO group $\hat{G}$ \cite{GODDARD19771}. The sum is taken over a Weyl chamber of the weight lattice $\Gamma^{\star}_{\hat{G}}$ of $\hat{G}$, while the classical factor reads
\begin{equation}
P_{G}(t;\vec{m}) = \prod_{j=1}^{q}\frac{1}{1-t^{d_{j}(\vec{m})}} \, ,    
\end{equation}
where $d_{j}(\vec{m})$ for $j=1,...,q$ are the degrees of the Casimir invariants of the residual gauge group. Also in this case let's consider the large $N$ limit of the Hilbert series (\ref{eq:hsgeneral}): we observe that only the term with $\vec{m}=(0,...,0)$ gives a contribution. Therefore we obtain
\begin{equation}
\lim_{N \to + \infty}\mathcal{C}\left[\mathscr{A}^N_{G}\right] = P_{G}(t^2;0,...,0)  = \textrm{PE}\left[\sum_{j=1}^{q}t^{2d_j}(\vec{0})\right] \, .   
\end{equation}
On the other hand, as explained in subsection \ref{halfBPS}, the half-BPS index of the $A-D-E$ theory precisely coincides with the counting of Casimir invariants in the corresponding group as described above. Hence, just like for the $A$-type, we could reproduce the $D$-type half-BPS index in \eqref{ID} by choosing $G$ above to be $\mathrm{O}(2k)$ to match \eqref{ID}. Thus, with the deconstruction of the type $D_k$ $\mathcal{N}=(2,0)$ theory in mind, we explore in the rest of this section the Coulomb branch Hilbert series of theories $\mathscr{A}^{N}_{G}$ for $G$ an orthogonal or a symplectic group, and use these results as a guide for finding the correct mirror theory, which will be of type $\mathscr{B}^{N}_{G_1,G_2}$ with (special)orthogonal and symplectic gauge groups. Discussion of how this relates to deconstruction is postponed until section \ref{secDeconstructionD}.

\begin{figure}[t]
    \centering
    \begin{tabular}{|c|c|c|c|c|}
    \hline 
      Type of O3   & Gauge algebra & S-dual & D5 splitting & NS5 splitting \\ \hline
       $O3^-$  &  $D_k$ & $O3^-$ & $B_k$ & $C_k$ \\
       $\tilde{O3}^-$  &  $B_k$ & $O3^+$ & $D_{k+1}$ & $C'_k$ \\
       $O3^+$  &  $C_k$ & $\tilde{O3}^-$ & $C'_k$ & $D_{k+1}$ \\
       $\tilde{O3}^+$  &  $C'_k$ & $\tilde{O3}^+$ & $C_k$ & $B_k$\\
         \hline
    \end{tabular}
    \caption{The first column enumerates all possible orientifold O3 planes. The second column then gives the gauge algebra of the theory with $k$ coinciding branes on the orientifold. The third column indicates how the orientifold planes transform under S-duality of type IIB string theory. The fourth and fifth columns give the gauge algebra of the world-volume theory between two half D5 or half NS5 branes after splitting on the corresponding orientifold. In particular, brane creations occur in some cases, as can be seen by the increase in rank of the gauge algebra. }
    \label{TableOrientifolds}
\end{figure}

\begin{figure}[t]
\begin{center}
    \begin{tabular}{|c|c|}
    \hline 
      $\mathfrak{g}$   & $\mathfrak{g}_1$, $\mathfrak{g}_2$ \\ \hline
       $D_k$  &  $D_k$, $C_k$ \\
       $B_k$  &  $D_{k+1}$, $C_k$ \\
       $C_k$  &  $B_k$, $C'_k$ \\
       $C'_k$  &  $B_k$, $C'_k$ \\
         \hline
    \end{tabular}
 \end{center}
    \caption{A theory with gauge algebra $\mathfrak{g}$, one adjoint hypermultiplet and $N$ flavors is mirror to a theory described by the circular quiver with $2N$ nodes with alternating gauge algebras $\mathfrak{g}_1$ and $\mathfrak{g}_2$ given by the table.  }
    \label{TableMirrorsAlg}
\end{figure}

Theories $\mathscr{A}^{N}_{G}$ with $G$ orthogonal or symplectic with rank $k$ can be realized in type IIB string theory as the world-volume theory of $k$ D3 branes on top of an O3 orientifold plane wrapping a circle, with $N$ additional D5 branes to account for the flavors. The precise group $G$ depends on the type of orientifold plane. The gauge \emph{algebra} of the theory on the world-volume of $k$ D3 branes on top of various kinds of O3 orientifold plane is recalled in Table \ref{TableOrientifolds}. In order to compute the mirror theory, we also need to keep track of brane creation phenomena when the D5 branes split into half-D5 when they coincide with the orientifold, and the relevant data are also summarized in Table \ref{TableOrientifolds}. It is then a simple exercise to compute the S-dual configuration, and to deduce the gauge \emph{algebra} of the mirror theories. The result is presented in Table \ref{TableMirrorsAlg}. 

In the following, we will see how mirror symmetry and consistency conditions allow to fix the gauge \emph{groups} in some of the mirror pairs. 

\subsection{\texorpdfstring{Mirror of the $\mathrm{O}(2k)$ theory}{}}
\label{SectionMirrorSymmetry}

The expression for the Coulomb branch Hilbert series of the first quiver can be deduced from the monopole formula \cite{Cremonesi:2013lqa}, using the additional property 
\begin{equation}
    P_{\mathrm{O}(2k)} (t ; \vec{m}) = P_{\mathrm{SO}(2k+1)} (t ; \vec{m}) \, . 
\end{equation}
and using for $\mathrm{O}(2k)$ the lattice of $\mathrm{SO}(2k+1)$, as explained in an appendix A of \cite{Cremonesi:2014uva}. 

Then the Coulomb branch Hilbert series for the theory $\mathscr{A}^N_{\mathrm{O}(2k)}$ of Figure \ref{figureTheoriesAB} reads 
\begin{equation}
\label{monopoleFormula}
    \mathcal{C}[\mathscr{A}_{N,\mathrm{O}(2k)}] = \sum\limits_{m_1 \geq \cdots \geq m_k \geq 0} t^{2N (m_1 + \cdots + m_k)} P_{\mathrm{O}(2k)} (t^2 ; m_1 , \cdots  , m_k) \, . 
\end{equation}
We can then evaluate, for instance,
\begin{eqnarray}
\label{CBHSkequals1}
    \mathcal{C}[\mathscr{A}_{N,\mathrm{O}(2)}] &=& \frac{1}{1-t^4} + \frac{1}{1-t^2} \sum\limits_{m \geq 1} t^{2Nm} \\
    &=& \frac{1+t^{2N+2}}{(1-t^4)(1-t^{2N})} = \mathrm{PE}\left[u^2+u^N+u^{N+1}-u^{2(N+1)}\right] \, , 
\end{eqnarray}
where we have introduced $u = t^2$. Note that this is the Hilbert series of a singularity $\mathbb{C}^2 / D_{N+2}$, or equivalently one rank zero instanton on this singular space. 
This allows us to deduce that the generic $k$ case is obtained by considering $k$ rank zero instantons on the same space, 
\begin{equation}
\label{mainCBHS}
   \textrm{PE} \left[ \nu \mathcal{C}[\mathscr{A}_{N,\mathrm{O}(2)}] \right] = 1+ \sum\limits_{k=1}^{\infty} \nu^k \mathcal{C}[\mathscr{A}_{N,\mathrm{O}(2k)}] \, . 
\end{equation}

Note that this is the same ``abelianization" as in \eqref{abelianization} but at finite $N$. As we will discuss in more detail below, this is analog to the $A$-type case, where the computation at hand is related to an instanton moduli space for a would-be instanton of rank zero. This formally corresponds to a pointlike instanton, whose only degrees of freedom are geometric. Hence the moduli space of $k$ of them is simply the $k$-fold symmetrized product of one. As we will discuss below, one way to understand this, analogously to the $A$ type case, is to rely on the brane picture. In the appropriate duality frame, there is a brane interpretation in terms of D2s and D6s on top of an orientifold O2 plane, and a similar instanton interpretation is possible. It is important to stress that this interpretation -- as well as the ``abelianization" property -- is lost had we considered $\mathrm{SO}(2k)$ instead of $\mathrm{O}(2k)$ (see the discussion around (\ref{SO4HS})). 

As an explicit check, \eqref{mainCBHS} predicts 
\begin{equation}
    \mathcal{C}[\mathscr{A}_{N,\mathrm{O}(4)}] = \frac{\mathcal{C}[\mathscr{A}_{N,\mathrm{O}(2)}](u)^2 + \mathcal{C}[\mathscr{A}_{N,\mathrm{O}(2)}](u^2)}{2} \, . 
\end{equation}
We can check this by directly evaluating (\ref{monopoleFormula}): 
\begin{equation}
    \mathcal{C}[\mathscr{A}_{N,\mathrm{O}(4)}] = \frac{u^{N+1}+u^{N+2}+u^{N+3}+u^{2 N+1}+u^{2 N+2}+u^{2
   N+3}+u^{3 N+4}+1}{(u-1)^2 (u+1)^2 \left(u^2+1\right)
   \left(u^N-1\right)^2 \left(u^N+1\right)} \, . 
\end{equation}

Taking the large $N$ limit, we have 
\begin{equation}
    \lim\limits_{N \rightarrow + \infty} \mathcal{C}[\mathscr{A}_{N,\mathrm{O}(2k)}] = P_{\mathrm{O}(2k)} (t^2 ; 0 , \cdots , 0) = \textrm{PE} \left[u^2 \frac{1-u^{2k}}{1-u^2} \right] \, . 
\end{equation}
As expected, this is precisely \eqref{ID} above.

So far we have worked in the ``magnetic theory". However, for our purposes we are more interested in the ``electric theory", which, appliying the standard rules, is $\mathscr{B}^N_{\mathrm{O}(2k),\mathrm{Sp}(k)}$. In order to check this, one can then compute the Higgs branch Hilbert series of the theory $\mathscr{B}^N_{\mathrm{O}(2k),\mathrm{Sp}(k)}$. However, this is more easily said than done, and one could think of at least three strategies to perform this calculation:
\begin{enumerate}
    \item One could try to enumerate operators using the letter counting method, then keeping only the gauge-invariant operators. 
    \item One could write down the superpotential of the theory, derive the F-terms and then use tools from algebraic geometry to compute the Hilbert series of the ideal generated by those F-terms. Then we have to integrate out the gauge degrees of freedom to obtain the Hilbert series $\mathcal{H}[\mathscr{B}^N_{\mathrm{O}(2k),\mathrm{Sp}(k)}]$ that counts chiral gauge-invariant operators on the Higgs branch. 
    \item Finally one could find another theory $\tilde{\mathscr{B}}^N_{\mathrm{O}(2k),\mathrm{Sp}(k)}$ whose Higgs branch would be identical to the one of $\mathscr{B}^N_{\mathrm{O}(2k),\mathrm{Sp}(k)}$, and then apply the previous methods on this other theory. 
\end{enumerate}
We will now explain the instructive reasons why the first two methods are unsuccessful in our case, and then show how the third strategy can be used to achieve our aims. 

Let us first explain why the Higgs branch Hilbert series of the theory $\mathscr{B}^{N}_{\mathrm{O}(2k),\mathrm{Sp}(k)}$ can not be computed using the standard strategy of letter-counting followed by a gauge integration. The argument is similar to the one outlined in section \ref{SectionComputationU}. For simplicity, let us focus on the theory $\mathscr{B}^{N}_{\mathrm{O}(2k),\mathrm{Sp}(k)}$ for $N=1$. The quiver is 
\begin{equation}
        \begin{tikzpicture}
\node[roundnode] (1) {$\mathrm{O}(2k)$};
\node[roundnode] (2) [right=of 1] {$\mathrm{Sp}(k)$};
\draw[-] (1) edge [bend left] (2);
\draw[-] (1) edge [bend right] (2);
\end{tikzpicture}  
    \end{equation}
and we call $H_1$ and $H_2$ the two half-hypermultiplets. We introduce the matrices 
\begin{equation}
    J = \left( 
    \begin{array}{cc}
        0 & -\mathbf{1}_k \\
        \mathbf{1}_k & 0
    \end{array}
    \right) \, , \qquad 
    M = \left( 
    \begin{array}{cc}
        0 & \mathbf{1}_k \\
        \mathbf{1}_k & 0
    \end{array}
    \right) \, . 
\end{equation}
The superpotential reads 
\begin{eqnarray}
    W &=& \mathrm{Tr} \left( H_1 J \phi_{\mathrm{Sp}(k)} J H_1^T M - H_1^T M \phi_{\mathrm{O}(2k)} M H_1 J \right. \\ \nonumber
    & & \left. + H_2 M \phi_{\mathrm{O}(2k)} M H_2^T J - H_2^T J   \phi_{\mathrm{Sp}(k)}  J H_2 M \right) \, ,
\end{eqnarray}
where $\phi_{\mathrm{Sp}(k)}$ and $\phi_{\mathrm{O}(2k)}$ are the scalars in the vector multiplets of the gauge group $\mathrm{Sp}(k)$ and $\mathrm{O}(2k)$ respectively. The relevant F-terms are 
\begin{eqnarray}
\label{FtermsSp}
    F_{\mathrm{Sp}(k)} &=& H_1^T M H_1 - H_2 M H_2^T \, , \\ 
    F_{\mathrm{O}(2k)} &=& H_2^T J H_2 - H_1 J H_1^T \, . 
\end{eqnarray}
In the ring quotiented by $F_{\mathrm{Sp}(k)}$, we can then compute 
\begin{equation}
  H_2 M F_{\mathrm{O}(2k)} M H_1 J = [H_1^T M H_1 J , H_2 M H_1 J] \, . 
\end{equation}
Therefore, we have the relation 
\begin{equation}
    \mathrm{Tr} \left( M H_1 J H_2 M F_{\mathrm{O}(2k)}  \right) = 0 \, . 
\end{equation}
This implies that the ideal defined by $F_{\mathrm{Sp}(k)}$ and $F_{\mathrm{O}(2k)}$ is not a complete intersection, and therefore we can not use letter counting. 

Having said that, one could envisage another method, namely compute the $\mathcal{F}^{\flat}$ Hilbert series using algebraic geometric tools (this is the method number 2 in the list above). However, this is doomed as well, for another reason: it is then impossible to perform the gauge integration over the disconnected group $\mathrm{O}(2k)$. Let us recall that for a class function $f$ defined on $\mathrm{O}(2k)$, we have 
\begin{equation}
\label{OnIntegration}
     \int_{\mathrm{O}(2k)} \d \eta_{\mathrm{O}(2k)}(X) f(X) = \frac{1}{2} \left[  \int \d \mu_{\mathrm{SO}(2k)} (z) f(z) + \int  \d \mu_{\mathrm{Sp}(k-1)} (z) f(z \mathcal{P} )  \right]  \, , 
\end{equation}
where $\d \eta_{G}$ is the Haar measure on the group $G$ and $\d \mu_{G}$ is the Haar measure on a maximal torus of $G$. Importantly, the operator $\mathcal{P}$ corresponds to a matrix of determinant $-1$ (see \cite{Bourget:2017tmt} for a more detailed explanation), and its appearance makes it crucial for the integrand to be written explicitly as a class function, i.e. a function that is invariant under conjugation by all the elements of the group. For connected groups, a character determines a class function in a unique way, but this is no longer true for disconnected groups. For instance, in the case of $\mathrm{O}(2)$, it is not possible to know whether a constant should be written as the character of the trivial representation, or as the character of the adjoint representation. But this distinction is crucial for our purposes, since $\int_{\mathrm{O}(2)} \d \eta_{\mathrm{O}(2)}(X) 1$ can then be interpreted in two different ways as
\begin{equation}
     \int_{\mathrm{O}(2)} \d \eta_{\mathrm{O}(2)}(X) \mathrm{Tr} \, \Phi_{\mathbf{1}} (X) = \frac{1}{2} \left[ \left( \oint_{\mid z \mid=1} \frac{\d z}{2 \pi i z} \right) + 1 \right] = 1 \, ,  
\end{equation}
where $\mathrm{Tr} \, \Phi_{\mathbf{1}} (X)$ denotes the character of the trivial representation. Alternatively it could also be interpreted as
\begin{equation}
     \int_{\mathrm{O}(2)} \d \eta_{\mathrm{O}(2)}(X) \mathrm{Tr} \, \Phi_{\mathbf{Adj}} (X) = \frac{1}{2} \left[ \left( \oint_{\mid z \mid=1} \frac{\d z}{2 \pi i z} \right) + (-1) \right] = 0 \, .   
\end{equation}

\begin{figure}[t]
    \centering
         \def\x{3}
\begin{tikzpicture}
\node[squarednode] (-1) at (-2*\x,0*\x) {$\mathrm{O}(1)$};
\node (0) at (0*\x,0*\x) {$2N$ nodes};
\node[roundnode] (1) at (1*\x,0*\x) {$\mathrm{Sp}(k)$};
\node[roundnode] (2) at (.707*\x,.707*\x) {$\mathrm{O}(2k)$};
\node[roundnode] (3) at (0*\x,1*\x) {$\mathrm{Sp}(k)$};
\node[roundnode] (4) at (-.707*\x,.707*\x) {$\mathrm{O}(2k)$};
\node[roundnode] (5) at (-1*\x,0*\x) {$\mathrm{Sp}(k)$};
\node[roundnode] (6) at (-.707*\x,-.707*\x) {$\mathrm{O}(2k)$};
\node[roundnode] (7) at (0*\x,-1*\x) {$\mathrm{Sp}(k)$};
\node[roundnode] (8) at (.707*\x,-.707*\x) {$\mathrm{O}(2k)$};
\draw[-] (-1) -- (5);
\draw[-] (1) -- (2);
\draw[-] (2) -- (3);
\draw[-] (4) -- (3);
\draw[-] (4) -- (5);
\draw[-] (6) -- (5);
\draw[-] (6) -- (7);
\draw[-] (8) -- (7);
\draw[-] (8) -- (1);
\end{tikzpicture}
    \caption{Quiver defining the $\tilde{\mathscr{B}}^N_{\mathrm{O}(2k),\mathrm{Sp}(k)}$ theory. }
    \label{FigBtildeTheoryO}
\end{figure}

We will now consider instead the quiver of Figure \ref{FigBtildeTheoryO}, which defines a theory $\tilde{\mathscr{B}}^N_{\mathrm{O}(2k),\mathrm{Sp}(k)}$. We claim that 
\begin{equation}
    \mathcal{H}[\mathscr{B}^{N}_{\mathrm{O}(2k),\mathrm{Sp}(k)}] = \mathcal{H}[\tilde{\mathscr{B}}^{N}_{\mathrm{O}(2k),\mathrm{Sp}(k)}] \, . 
\end{equation}
The argument is similar to the simpler one presented in section \ref{SectionComputationU} for the theories with gauge groups $\mathrm{U}(k)$. For simplicity, let us consider again the case $N=1$. The superpotential has an additional contribution $\mathrm{Tr} (Z^T J \phi_{\mathrm{Sp}(k)} J Z)$ where $Z$ is the half-hypermultiplet which transforms under the global $\mathrm{O}(1)$. The F-terms (\ref{FtermsSp}) are modified accordingly to 
\begin{equation}
    F_{\mathrm{Sp}(k)} = H_1^T M H_1 - H_2 M H_2^T + Z Z^T \, ,  
\end{equation}
and we can eliminate $Z$ exactly like in (\ref{ZZelimination}). Note that this would not be possible if we added an $\mathrm{Sp}(1)$ flavor node on an $\mathrm{O}(2k)$ gauge node, since then the first equality in (\ref{ZZelimination}) can not hold (the left-hand side in now a $2 \times 2$ matrix). 

Thus, all in all, just like in the unitary case, the $\mathrm{O}(1)$ global symmetry that has been added to one of the $\mathrm{Sp}(k)$ is crucial, since it allows to use letter counting and performing explicitly the $\mathrm{O}(k)$ integration. While we have mostly presented the auxiliary theory as a mathematical trick to compute the Higgs branch Hilbert series, one may find a physical argument along the lines of the unitary case, in fact putting in firmer grounds the instanton analogy behind the ``abelianization" of the Higgs branch. However, this case is much more involved. Note first that, upon adding the $\mathrm{O}(1)$ symmetry, the $\mathrm{Sp}(k)$ node has an odd number of flavors, and hence needs a half-integer Chern-Simons in order to cancel a parity anomaly. The Chern-Simons term breaks half of the supersymmetry and the theory becomes $\mathcal{N}=2$. Moreover, in addition to the fields in the quiver and the superpotential, we also have monopole operators $T$, $\tilde{T}$ satisfying a quantum relation of the form $\tilde{T}T=\phi_{\mathrm{Sp}(k)}^N$. However, since we are interested in the ``instanton branch" of the theory where adjoint scalars are set to zero, we can as well consistently set to zero the monopole operators. Therefore, the naive classical computation of the Higgs branch for the flavored theory gives the correct computation for the ``instanton branch" in the honest $\mathcal{N}=2$ theory.  This is very similar to the case of instantons on $\mathbb{C}P^2$ \cite{Mekareeya:2014kca,Pini:2015lka}. To gain further intution, the extended theory can be embedded into string theory. To simplify the discussion, let us just consider the unorbifolded case, namely the $\mathrm{Sp}(k)$ theory with a symmetric hypermultiplet, which can be constructed in IIA on $k$ $D2$ on top of an $O2^+$ plane. Adding one flavor to this amounts to adding a half $D6$ brane. This turns the orientifold into an $\widetilde{O2}^+$, which requires the Romans' mass to be odd. In turn, an odd Romans' mass induces a half-integer CS on the $D2$ world-volume.\footnote{An indirect argument is to consider an $O2^-$ with a stuck half-NS brane, which turns it into a $O2^+$ to the other side. Turning on a Romans' mass induces a tadpole in the NS world-volume, which has to be cancelled by having an extra half-D6 brane ending on it. This shows that the $\mathrm{Sp}(2k)$ theory with a symmetric and an odd number of flavors must have a half-integral CS. We thank Oren Bergman for insightful conversations on this point.} At this point, we see that, independently on the Romans' mass, the Higgs branch of the theory corresponding to bound states of $D2-D6$ counts geometric degrees of freedom in very much the same spirit as in the unitary case, thus providing as well a heuristic motivation for the auxiliary theory as a tool to compute the desired Higgs branch Hilbert series. Moreover, it also explains the ``abelianization" property, since we are considering truly pointlike instantons, in such a way that the ensemble of $k$ of them simply corresponds to the $k$-fold symmetrized product of the one-instanton case.

We now proceed to the computation of the Higgs branch Hilbert series of $\tilde{\mathscr{B}}^N_{\mathrm{O}(2k), \mathrm{Sp}(k)}$. We note that the only difference with the Hilbert series of the $\mathscr{B}^N_{\mathrm{O}(2k), \mathrm{Sp}(k)}$ theory is the presence of the following further term in the integrand
\begin{equation}
    \textrm{PE}[tw\chi_{\mathrm{Sp}(k)}] \, ,
\end{equation}
where $w$ is the fugacity of the $\mathrm{O}(1)$ global symmetry group, while $\chi_{\mathrm{Sp}(k)}$ are the characters of the fundamental representation of the $\mathrm{Sp}(k)$ gauge group. This in fact provides an additional check on the validity of our method: the end result after gauge integration must not depend on $w$. Indeed, we find in all our calculations that although the integrand depends on $w$, the results after gauge integration involve only $w^2=1$.

\subsubsection*{Computation}
\label{sectionHBHSBtilde}

The Higgs branch Hilbert series of the theory $\tilde{\mathscr{B}}^N_{\mathrm{O}(2k), \mathrm{Sp}(k)}$ can be computed using letter counting, followed by gauge integration: 
\begin{equation}
    \mathcal{H} [\tilde{\mathscr{B}}^N_{\mathrm{O}(2k), \mathrm{Sp}(k)}] = \int \left( \prod\limits_{i \in \mathbb{Z}_N}  \d \eta_{\mathrm{O}(2k)}(X_i) \d \eta_{\mathrm{Sp}(k)}(Y_i) \right) \mathcal{F}^{\flat} (X_1 , \dots , X_N , Y_1 , \dots , Y_N) \, . 
\end{equation}
The integrand $\mathcal{F}^{\flat}$ is a class function, so we can use the Weyl integration formula (\ref{OnIntegration}), 
\begin{eqnarray}
\label{bigSum}
    \mathcal{H} [\tilde{\mathscr{B}}^N_{\mathrm{O}(2k), \mathrm{Sp}(k)}] &=& \\ 
    & & \kern-5em 2^{-N} \sum\limits_{\epsilon_1 , \dots , \epsilon_N = \pm 1} \int \left( \prod\limits_{i \in \mathbb{Z}_N}  \d \mu_{\mathrm{O}(2k)}^{\epsilon_i} (x_i) \d \mu_{\mathrm{Sp}(k)}(y_i) \right) \mathcal{F}^{\flat} (x_1 , \dots , x_N , y_1 , \dots , y_N) \, , \nonumber 
\end{eqnarray}
where we have introduced the notation $\d \mu_{\mathrm{O}(2k)}^{+1} = \d \mu_{\mathrm{SO}(2k)}$ and $\d \mu_{\mathrm{O}(2k)}^{-1} = \d \mu_{\mathrm{Sp}(k-1)}$. The summand where $\epsilon_1 = ... = \epsilon_N = 1$ corresponds to the Higgs branch Hilbert series of the $\tilde{\mathscr{B}}^N_{\mathrm{SO}(2k), \mathrm{Sp}(k)}$. 

Let us illustrate this computation on an example. We take $N=4$ and $k=1$. The integrals in (\ref{bigSum}) then depend only on the number of plus and minus signs in $(\epsilon_1 ,..., \epsilon_4)$, and they are evaluated to 
\begin{eqnarray}
   (+,+,+,+)&\rightarrow&  \nonumber I_4=\frac{1+3t^2+3t^4+t^6+15t^8-3t^{10}-3t^{12}-t^{14}}{1-t^2-t^8+t^{10}} \, , \\ \nonumber
   (+,+,+,-) &\rightarrow& I_3=\left(1+t^2\right)^2 \, ,\\ \nonumber
   (+,+,-,-) &\rightarrow& I_2=1-t^4 \, ,\\
    (+,-,-,-) &\rightarrow&  I_1=\left(1-t^2\right)^2 \, ,\\ \nonumber
   (-,-,-,-)&\rightarrow& I_0=\frac{\left(1-t^2\right)^3}{1+t^2} \, . \nonumber
    \end{eqnarray}
One note in passing that the Higgs branch Hilbert series of the $\tilde{\mathscr{B}}^N_{\mathrm{SO}(2k), \mathrm{Sp}(k)}$ theory does not have a palindromic numerator. Taking the appropriate linear combination (\ref{bigSum}) of these partial results, one obtains 
\begin{eqnarray}
    \mathcal{H} [\tilde{\mathscr{B}}^{N=4}_{\mathrm{O}(2), \mathrm{Sp}(1)}] &=& \frac{1}{16} (I_4 + 4I_3 + 6I_2 + 4I_1 + I_0) \\
    &=& \frac{1-t^2+t^4-t^6+t^8}{(1-t)^2 (1+t)^2 \left(1+t^2\right) \left(1+t^4\right)} = \mathrm{PE}[t^4 + t^8 + t^{10} - t^{20}] \, . \nonumber
\end{eqnarray}
One finds perfect matching with the Coulomb branch computation (\ref{mainCBHS}). 

On a large series of examples, the conclusion is that 
\begin{equation}
     \mathcal{H}[\mathscr{B}^N_{\mathrm{O}(2k),\mathrm{Sp}(k)}]  = \mathcal{C}[\mathscr{A}^N_{\mathrm{O}(2k)}] \, . 
\end{equation}

In the remainder of this section, we use our techniques to investigate other mirror pairs involving orthogonal and symplectic gauge groups. The reader most interested in deconstruction can go directly to section \ref{secDeconstructionD}. 

\subsection{\texorpdfstring{Mirror of the $\mathrm{SO}(2k)$ theory}{}}

For completeness, and also to have a comparison point with the previous subsection, we now identify the mirror of the $\mathscr{A}^N_{\mathrm{SO}(2k)}$ theory. This is obtained from the $\mathscr{A}^N_{\mathrm{O}(2k)}$ by ungauging the $\mathbb{Z}_2$ subgroup, and as a consequence the mirror theory will similarly be obtained by ungauging a $\mathbb{Z}_2$. 

The Coulomb branch Hilbert series can be computed from the monopole formula. One obtains, for instance, for $k=1$
\begin{equation}
\label{CBSO2}
    \mathcal{C} [\mathscr{A}^N_{\mathrm{SO}(2)}] =  \frac{1+t^{2N}}{\left(1-t^2\right) \left(1-t^{2 N}\right)} \, . 
\end{equation}
As expected, because of the accidental equality $\mathrm{SO}(2)=\mathrm{U}(1)$, this is the moduli space of one rank zero instanton on $\mathbb{C}^2/A_{N-1}$. But when the gauge group is $\mathrm{SO}(2k)$, we don't expect the Hilbert series to be the moduli space of $k$ particles on this space. Consistently with the remark in the paragraph after (\ref{mainCBHS}), there is no obvious brane interpretation for this theory. Indeed, for $k=2$ we have
    \begin{equation}
    \label{SO4HS}
\mathcal{C} [\mathscr{A}^N_{\mathrm{SO}(4)}] = \frac{1+2t^{2N+2}+t^{2N+4}+t^{4N}+2t^{4N+2}+t^{6N+4}}{\left(1-t^4\right)^2 \left(1-t^{2N}\right)^2
   \left(1+t^{2 N}\right)}\, , 
    \end{equation} 
and we don't have a relation analogous to (\ref{mainCBHS}). 

The Higgs branch Hilbert series computation follows the logic of section \ref{sectionHBHSBtilde}. The $\mathbb{Z}_2$ ungauging is then performed by fixing one $\epsilon$, for instance $\epsilon_N$, to the value $1$ and summing over the remaining $\epsilon_1 , \dots , \epsilon_{N-1}$ in equation (\ref{bigSum}). For $N=1$, this clearly reduces to taking a circular quiver with two nodes $\mathrm{SO}(2k)$ and $\mathrm{Sp}(k)$. For higher values of $N$, this computes the Higgs branch of a circular quiver theory with one gauge node $\mathrm{SO}(2k)$, $N-1$ nodes $\mathrm{O}(2k)$ and $N$ nodes $\mathrm{Sp}(k)$. 

Let us take again the example $N=4$ and $k=1$ from the previous paragraph. We obtain 
\begin{eqnarray}
    \mathcal{C} [\mathscr{A}^{N=4}_{\mathrm{SO}(2)}] &=& \frac{1}{8} \left(I_4 + 3I_3+3I_2+I_1\right) \\
    &=& \frac{1+t^8}{1-t^2-t^8+t^{10}} = \mathrm{PE}\left[t^2+2 t^8-t^{16}\right] \, , \nonumber
\end{eqnarray}
in agreement with (\ref{CBSO2}).

\subsection{\texorpdfstring{Other theories}{}}

In this section, we comment briefly on theories $\mathscr{A}^N_{G}$ with $G$ of type B and C. 

By construction, the Coulomb branch Hilbert series for the theory $\mathscr{A}^N_{\mathrm{Sp}(k)}$ is equal to the Coulomb branch Hilbert series of the $\mathscr{A}^N_{\mathrm{O}(2k)}$ theory (\ref{mainCBHS}). Indeed the P-factors and the GNO lattices of magnetic charges in the summation of the monopole formula are identical in all cases. Because of the data of Table \ref{TableMirrorsAlg}, we expect the mirror theory to be a circular quiver with gauge algebras $B_k$ and $C_k$. We can check this against a Higgs branch Hilbert series. It appears that it is not possible to determine using this method if the gauge group of type $B_k$ is $\mathrm{SO}(2k+1)$ or $\mathrm{O}(2k+1)$, as remarkably, both theories turn out to have the same Higgs branch Hilbert series, which is equal to the Coulomb branch Hilbert series of $\mathscr{A}^N_{\mathrm{Sp}(k)}$. 

Finally, the mirror of the theory $\mathscr{A}^N_{G}$ with $G = \mathrm{O}(2k+1)$ or $G=\mathrm{SO}(2k+1)$ is more difficult to apprehend. While the Coulomb branch of $\mathscr{A}^N_{\mathrm{SO}(2k+1)}$ is characterized, for the same reasons as before, by (\ref{mainCBHS}), Table \ref{TableMirrorsAlg} suggests that the mirror theory should be described by a circular quiver with gauge algebras $D_{k+1}$ and $C_k$. However, a naive counting of the number of hypermultiplet for the $D_{k+1}$ gauge nodes indicates that the theory would be bad, in the sense of \cite{Gaiotto:2008ak}, making the computation of the Coulomb branch more tricky \cite{Assel:2017jgo}, and moreover, the classical analysis of the moduli space along the lines of \cite{Argyres:1996hc} suggests that there is no pure Higgs branch: the effective number of flavors for each $D_{k+1}$ gauge group is $n_f = 2k$, to be compared to $n_c = 2k+2$ colors. It would be interesting to clarify the structure of the moduli space of these theories, both with $\mathrm{O}(2k+2)$ and $\mathrm{SO}(2k+2)$ gauge groups.

\section{\texorpdfstring{Towards deconstruction of the type D theory}{}}
\label{secDeconstructionD}

%Elaborating on the deconstruction of the $A_k$-type (2,0) 6d theory of \cite{ArkaniHamed:2001ie}, we have argued that certain quantities -- such as the half-BPS index of the theory-- can be computed upon further compactification in an $S^1$ through a 3d theory. This suggested a generalization to the $D_k$-type (2,0) 6d theory which reproduces the expected half-BPS index as we have shown. 

We have argued that the half-BPS index of the type $D_k$ $(2,0)$ theory can be obtained by the Higgs branch Hilbert series of the $\mathscr{B}_{\mathrm{O}(2k),\mathrm{Sp}(k)}^N$ theory. It is then natural to wonder to what extent our 3d theory -- or its 4d uplift -- deconstructs the 6d (2,0) theory of type $D_k$.

Recall that, as argued in \cite{ArkaniHamed:2001ie}, the $A_k$ type 6d (2,0) theory is deconstructed by a certain scaling limit in the Higgs branch of the 4d $\mathcal{N}=2$ $N$-noded necklace quiver theory with $\mathrm{U}(k)$ gauge groups, where in particular $N$ is taken to infinity. An illuminating point of view can be obtained by $T$-dualizing along the Hopf fiber-like direction of the orbifold. One then finds a IIA configuration with $k$ D4 branes along $(1236)$ and $N$ NS5 branes along $(12345)$, sitting at fixed points in the circle parametrized by $x^6$. Then, moving into the Higgs branch amounts to a recombination of all the D4 pieces between each pair of NS into $k$ D4's, which can then be moved away from the NS in a transverse direction. This leaves us with $k$ D4 branes wrapping $\mathbb{R}^4\times S^1$ in flat space, whose world-volume description is in terms of the maximally SUSY theory with $\mathrm{U}(k)$ gauge group on $\mathbb{R}^4\times S^1$. In turn, the instanton spectrum of this theory is expected to be precisely the KK tower which allows to identify the theory with the 6d $A_k$ theory on $\mathbb{R}^4\times \mathbb{T}^2$. It is in this sense that the 4d quiver theory, in the appropriate limit, deconstructs the 6d theory.

In view of this, it is tempting to wonder to what extent this picture can be extended to the $D_k$-type 6d theory. The natural guess would be to add an $O4^-$ parallel to the $k$ D4-branes in the IIA picture. More explicitly, let us consider $k$ D4 branes on top of an $O4^-$ plane along $(x^1,x^2,x^3,x^6)$ and $N$ NS5 branes along $(x^1,x^2,x^3,x^4,x^5)$. If separated from the NS5's in the transverse space, one would be left with $k$ D4 branes plus an $O4^-$, whose uplift is in terms of $k$ M5 branes and $OM5$ and whose low energy description is the desired $D_k$ 6d theory (on $\mathbb{R}^4\times \mathbb{T}^2$).

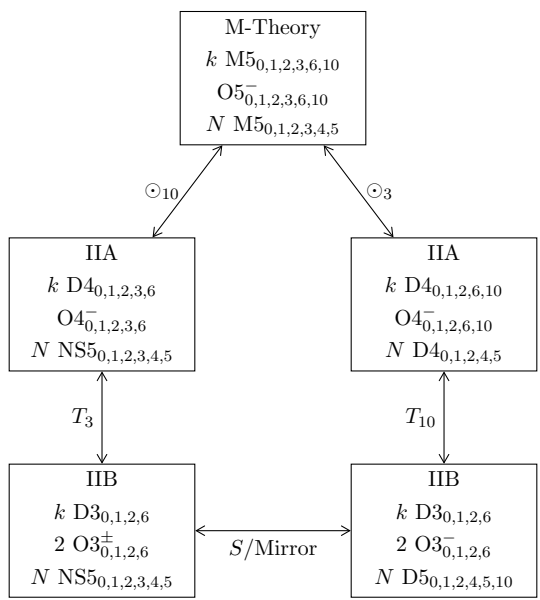
\begin{figure}[t]
    \centering
    \def\x{2}
\begin{tikzpicture}[scale=0.75,every node/.style={scale=0.75}]
\node[draw,text width=3cm,align=center] (1) at (4*\x,4*\x) {M-Theory \\ $k$ M5$_{0,1,2,3,6,10}$ \\ O5$^-_{0,1,2,3,6,10}$ \\ $N$ M5$_{0,1,2,3,4,5}$};
\node[draw,text width=3cm,align=center] (2) at (2.5*\x,2*\x) {IIA \\ $k$ D4$_{0,1,2,3,6}$ \\ O4$^-_{0,1,2,3,6}$ \\ $N$ NS5$_{0,1,2,3,4,5}$};
\node[draw,text width=3cm,align=center] (3) at (5.5*\x,2*\x) {IIA \\ $k$ D4$_{0,1,2,6,10}$ \\ O4$^-_{0,1,2,6,10}$ \\ $N$ D4$_{0,1,2,4,5}$};
\node[draw,text width=3cm,align=center] (6) at (2.5*\x,0*\x) {IIB \\ $k$ D3$_{0,1,2,6}$ \\ $2$ O3$^{\pm}_{0,1,2,6}$ \\ $N$ NS5$_{0,1,2,3,4,5}$};
\node[draw,text width=3cm,align=center] (7) at (5.5*\x,0*\x) {IIB \\ $k$ D3$_{0,1,2,6}$ \\ $2$ O3$^-_{0,1,2,6}$ \\ $N$ D5$_{0,1,2,4,5,10}$};
\draw [arrow] (1) -- (2) node[midway,left] {$\odot_{10}$};
\draw [arrow] (1) -- (3) node[midway,right] {$\odot_{3}$};
\draw [arrow] (2) -- (6) node[midway,left] {$T_3$};
\draw [arrow] (3) -- (7) node[midway,left] {$T_{10}$};
\draw [arrow] (6) -- (7) node[midway,below] {$S$/Mirror};
\end{tikzpicture}
    \caption{Summary of the dualities for deconstruction of type $D$. Directions 3, 6 and 10 are compact $S^1$, while the other directions are non-compact. }
    \label{summaryDualitiesOrth}
\end{figure}

Very much like in the $A$-type above, we would like to think of the separation as motion along the Higgs branch of the theory arising when the NS are brought to coincide with the $k$ D4+ $O4^-$. However, when that happens the situation is now much richer. To begin with, the $N$ NS fractionate into $2N$ half-NS. Then, the $O4$ changes from $O4^+$ into $O4^-$ and vice-versa when crossing each half-NS. Thus we end up with $2N$ segments among the half-NS which alternate $k$ D4+$O4^-$ with $k$ D4+$O4^+$. Reassuringly, this precisely corresponds to the $[\mathrm{O}(2k)\times \mathrm{Sp}(k)]^N$ necklace theory $\mathscr{B}_{\mathrm{O}(2k),\mathrm{Sp}(k)}^N$ considered above. 

We can further T-dualize this to obtain the 3d version, which is in terms of $2k$ half-D3 branes on top of an $O3^{\pm}$ along $(x^1,x^2,x^6)$ , an $2N$ half-NS5 branes along $(x^1,x^2,x^3,x^4,x^5)$. This configuration is nothing but the orientifold version of the type $A$ case above. Thus, similar arguments to those above lead to the mirror configuration with $2k$ half-D3 branes on top of an $O3^{-}$ along $(x^1,x^2,x^6)$ , an $2N$ half-D5 branes along $(x^1,x^2,x^3,x^4,x^5)$. Further T-duality along $x^6$ gives $2k$ D2 on top of an $O2^-$ and $2N$ half-D6 branes, from which the mirror $\mathrm{O}(2k)$ theory with $\mathrm{Sp}(2N)$ flavor symmetry, which we dubbed $\mathscr{A}_O^N$, can be easily read. We summarize the situation in Figure \ref{summaryDualitiesOrth}.

\begin{figure}[t]
    \centering
    \includegraphics[width=.45\paperwidth]{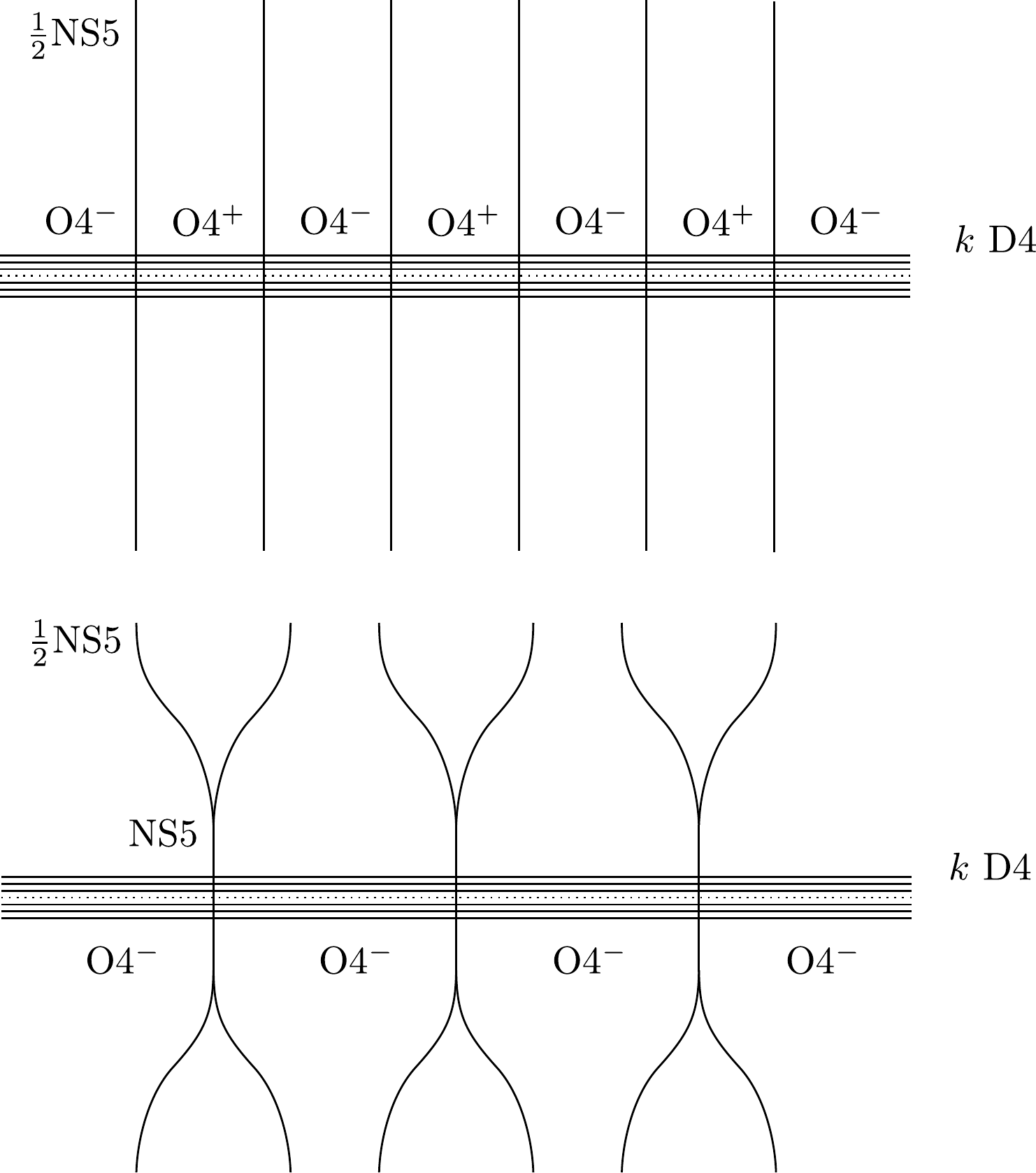}
    \caption{The upper part of the Figure shows the naive brane configuration that reproduces at low energies the quiver theory $\mathscr{B}^N_{\mathrm{O}(2k),\mathrm{Sp}(k)}$, which is the type IIA configuration on the left of Figure \ref{summaryDualitiesOrth}. The horizontal axis is the direction 6, and the vertical axis is the direction 5. Periodic identification between the left part and the right part is assumed. As a result of the bending described in the text, a more realistic view is pictured in the lower part, where the half-NS5 branes recombine to give full NS5, and the segments of O4$^+$ shrink to vanishing size. }
    \label{figBranesBending}
\end{figure}

Note that the converse process of the Higgsing which detaches, in the IIA set-up, the $k$ D4+$O4^-$ from the $2N$ half-NS5's is much less clear in the $D$ type case, since, to begin with, being the NS fractional, it seems that they cannot be moved away from the orientifold. An intimately related issue is that the $\mathscr{B}_{\mathrm{O}(2k),\mathrm{Sp}(k)}^N$ 4d quiver theory is not conformal.\footnote{The $\mathrm{U}(k)$ necklace theory is non-conformal either, yet in that case in a very mild sense: the $\mathrm{U}(1)$ are IR free and decouple in the IR, leaving us with the $\mathrm{SU}(k)^N$ theory which is conformal.} In fact, a short computation shows that the $\beta$ function for each type of node is
\begin{equation}
\beta_{\mathrm{O}}=-4\,,\qquad \beta_{\mathrm{Sp}}=4\, .
\end{equation}
Thus, while the total $\beta$ function vanishes, the $\mathrm{O}$ nodes flow to weak coupling in the IR while the $\mathrm{Sp}$ nodes flow to strong coupling. Here we wish to suggest a scenario which could dynamically provide a deconstruction of the $D_k$ theory. The key observation is that the $\mathrm{Sp}$ nodes flow to strong coupling, which, in the IIA scenario described above, implies a bending of the NS branes in such a way that the segments bewteen half-NS bounding $D4-O4^+$ become of vanishing size (see Figure \ref{figBranesBending}). Thus, restricting to the origin of the Coulomb branch, we may effectively think that each pair of half-NS bounding $D4-O4^+$ have recombined into full NS. Therefore, the remaining $N$ $D4-O4^-$ segments can be recombined and moved out of the NS (or alternatively, the NS, now physical NS, can be moved out of the orientifold). Thus, this would provide a dynamical mechanism by which the left-over theory, in the Higgs branch, is the maximally SUSY 5d theory on $\mathbb{R}^4\times S^1$ with orthogonal gauge group, which, on general grounds, just as before should be equivalent to the $D_k$ theory on $\mathbb{R}^4\times \mathbb{T}^2$.

\section{Conclusion and future directions}
\label{conclusions}

In this paper we have studied the deconstruction of the type $D$ 6d $(2,0)$ theory. Our main tool has been the half-BPS limit of the 6d index. This is a particularly simple observable which counts operators which can be traced from 6d to 4d when we consider the 6d theory on $\mathbb{R}^4\times \mathbb{T}^2$. In the 4d deconstructing theory the relevant operators to count are chiral operators in the Higgs branch. The crucial observation is that such counting can be equally done upon reduction to 3d, which would amount to the 6d theory on $\mathbb{R}^4\times \mathbb{T}^3$. Using mirror symmetry we can likewise compute the half-BPS index by counting dressed monopole operators in the Coulomb branch of the ``magnetic" theory. This computation is in a sense much simpler than the ``electric" version, since while the latter involves an integration over the gauge group projecting to gauge-singlets, the Coulomb branch formula is simply a sum. Because of this it is very easy to take the deconstruction limit -- which in this language simply amounts to taking a large number of nodes in the ``electric" theory. On general grounds, this way one recovers a counting of the Cartans of the group, which is precisely the structure of the half-BPS index in 6d. This automatically ensures the matching of the index for the type $A$ theory and, at the same time, suggests a candidate mirror to a theory deconstructing the type $D$ theory. To be precise, we find that a $\mathrm{O}(2k)$ theory with an adjoint hypermultiplet and $2N$ vector half-hypermultiplets reproduces on its Coulomb branch, in the large $N$ limit, the half-BPS index of the 6d theory. 

The next step towards the deconstruction of the 6d type $D$ theory is to consider the mirror to the magnetic theory, whose Higgs branch will reproduce the 6d half-BPS index. Such ``electric" theory, uplifted to 4d, would be the natural candidate for a theory deconstructing the 6d type $D$ $(2,0)$ theory on $\mathbb{R}^4\times \mathbb{T}^2$. In the case at hand, we find that such electric theory is a $[\mathrm{O}(2k)\times \mathrm{Sp}(k)]^N$ circular quiver. As discussed, this not only follows from the brane configuration engineering the system, but can also be supported by the computation of the Higgs branch Hilbert series. In order to show this matching we introduced an auxiliary theory with an enlarged matter sector. As argued, because of the particular extra matter content chosen, the auxiliary theory has the same Higgs branch as the original one, with the extra bonus that one can use the much simpler technique of letter counting. This allowed us to explicitly show that the Higgs branch of the $[\mathrm{O}(2k)\times \mathrm{Sp}(k)]^N$ circular quiver reproduces the 6d half-BPS index.

The way to arrive to the $[\mathrm{O}(2k)\times \mathrm{Sp}(k)]^N$ circular quiver as deconstructing theory for the 6d $D$-type theory did not rely on any string construction, and was only based on the requirement that the 6d half-BPS index must be reproduced. However, the resulting circular quiver theory can be engineered on a stack of $k$ D4 branes on top of an $O4$ plane on a circle with $2N$ half-NS5 branes. Amusingly, this is precisely what one would have naively guessed as deconstructing theory for the 6d type $D$ case, since it corresponds to the same set-up as for the type $A$ case only with the addition of an orientifold. Thus, the emerging picture would seem to be consistent. 

Nevertheless, an important point in the deconstruction programe is that it should be possible to show that, upon going to the (equal VEV) Higgs branch, the theory becomes a discretized version of the maximally SUSY 5d theory with $D$ gauge group as in \cite{Lambert:2012qy}. In the case at hand this is by no means obvious. This can be easily argued in the brane picture, where, because of having half-NS branes, detaching the $k$ D4's together with the $O4$ seems impossible. A related issue is that, as opposed to the usual deconstruction procedure, in this case the deconstructing theory is not conformal. However, an observation is that the beta-functions for each node are such that the $\mathrm{Sp}$ groups hit infinite coupling at a point along the RG flow such that the $\mathrm{O}$ groups remain at finite coupling. While it is very hard to analyze the gauge theory dynamics, we may turn to the brane picture, where such infinite coupling for the $\mathrm{Sp}$ nodes stands for a bending of the half-NS5 branes such that they meet at some point as in Figure \ref{figBranesBending}. Note that, because of the orientifold, the collision of the NS5's seems unavoidable. While it is not known how to describe such collision in string theory, we may conjecture that in the end of the day the two half-NS will merge into a full NS. Thus, when all the branes are one on top of the other and on top of the orientifold -- which amounts to the origin of the Coulomb branch --, close to the 4-branes the NS5 look like $N$ full NS5, so that the detaching of the stack of $k$ D4's plus the $O4^-$ plane at the same time is possible. Note as well that this dynamically chooses $O4^-$ rather than $O4^+$ -- in other words, the maximally SUSY 5d theory with $\mathrm{O}(2k)$ group rather than $\mathrm{Sp}(k)$ emerges naturally from the dynamics. This is however highly conjectural, and it would be crucial to analyze this problem in detail, some of whose aspects are in fact interesting \textit{per se} in string theory -- such as what happens when the two half-NS5 collide, or alternatively, what is the IR dynamics of the $[\mathrm{O}(2k)\times \mathrm{Sp}(k)]^N$ circular quiver. 

It is interesting to note that, as a 3d theory, the $[\mathrm{O}(2k)\times \mathrm{Sp}(k)]^N$ circular quiver is a bad theory in the sense of \cite{Gaiotto:2008ak}. This is likely related to the subtleties raised above. However, as shown in \cite{Assel:2017jgo}, typically bad theories just correspond to theories for which there is a non-trivial RG flow along which the monopole operators which would seem to have R-symmetries below the unitary bound decouple. It would be very interesting to analyze these aspects in this particular case. Note however that for the circular quivers we are interested on the Higgs branch, which should be insensitive to this problem.

%We have proposed a candidate theory for the deconstruction of the type D $(2,0)$ theory, and verified that it passes the simplest non-trivial consistency test. However, one might want to go further and proceed as in \cite{Hayling:2017cva} with the comparison of full partition functions on $\epsilon$-deformed spheres. This will be addressed in a subsequent paper (? XXX). Moreover, fully exploring the deconstruction potential and the explicit limits to be taken is an exciting and certainly fruitful endeavour which we aim to take up. 

This work also opened a window on three-dimensional mirror symmetry for a large class of theories, some of which are still fairly mysterious, like the $\mathscr{A}^N_{\mathrm{O}(2k+1)}$ theory. It would be interesting to extend some of the tools put at work in this paper, combining algebraic techniques to physical insight, to understand the full class of theories. From the string theory point of view, this boils down in parts to determining whether the gauge group on the world-volume of branes on top of the appropriate orientifold plane is of type $\mathrm{O}$ or $\mathrm{SO}$, a question that might involve non-local physics, and which has been more and more actively investigated in recent years \cite{Cabrera:2017ucb}.

\section*{Acknowledgements}

We would like to thank O. Bergman,  A. Hanany, J. Hayling, C. Papageorgakis and E.Pomoni for very insightful comments and collaborations. D.R-G and A.B. acknowledge support from the EU CIG grant UE-14-GT5LD2013-618459, the Asturias Government grant FC-15-GRUPIN14-108 and Spanish Government grant MINECO-16-FPA2015-63667-P. D.R-G is partly supported by the Ramon y Cajal grant RYC-2011-07593. A.P. is supported by the German Research Foundation (DFG) via the Emmy
Noether program “Exact results in Gauge theories".

\begin{appendix}
\section{Complete Intersections and Hilbert Series}
\label{A1}

\subsection{Complete intersections and regular sequences}
Consider the polynomial ring $R = \mathbb{C}[x_1 , \dots x_n]$ in $n$ variables. 
We say that a sequence of non-constant polynomials $P_1 , P_2 , \dots , P_r$ is a regular sequence if for all $i=1,\dots , r$, $P_i$ is not a zero divisor modulo the partial ideal $(P_1 , \dots , P_{i-1})$.\footnote{One also usually requires $(P_1 , P_2 , \dots , P_r)R \neq R$, but we will not make use of this. } 
Let $I$ be an ideal of $R$. Then we have the following theorem \cite{stanley1978hilbert}: the ring $R/I$ is a complete intersection if and only if $I$ is generated by a regular sequence of homogeneous polynomials $P_1 , P_2 , \dots , P_r$ of degrees $d_1 , \dots , d_r$. In this case, the Hilbert series of $R/I$ is given by
\begin{equation}
    H_{R/I} (t) = \frac{\prod\limits_{i=1}^r (1-t^{d_i})}{(1-t)^n} \, . 
\end{equation}
In more down-to-earth physical terms, this means that we can use the ``letter-counting" technique to write down the Hilbert series.

\subsection{A counter example of complete intersection}
\label{secExNCI}

Let us now give an important example of an ideal which does not define a complete intersection. Consider the ring in $2 N^2$ variables $R = \mathbb{C}[x_{ij},y_{ij}]$ with $i,j = 1 \dots , N$, and the ideal $I$ defined by 
\begin{equation}
    XY-YX=0 \, , \qquad X = (x_{ij})_{i,j=1 \dots , N} \, , \qquad Y = (y_{ij})_{i,j=1 \dots , N} \, . 
\end{equation}
The ideal $I$ is generated by the polynomials $P_{ij}$ which are the matrix elements of $P = XY-YX$. There are $N^2-1$ independent such polynomials, because the trace of $XY-YX$ vanishes, so we just redefine $P_{N,N} = - P_{1,1} - \dots - P_{N-1 , N-1}$. But these polynomial satisfy another relation, 
\begin{equation}
\label{zeroDiv}
    \mathrm{Tr} \left( X P \right) = 0 \, . 
\end{equation}
Let us now consider the ideal $I'$ defined by 
\begin{equation}
    I' = (P_{ij})_{i,j=1 , \dots , N \textrm{ and } (i,j) \neq (N,N) , (i,j) \neq (N-1 , N)} \, , 
\end{equation}
and rewrite (\ref{zeroDiv}) in the ring $R/I'$: 
\begin{equation}
    x_{N,N-1} P_{N-1,N} = 0 \textrm{ modulo } I' \, . 
\end{equation}
This means that $P_{N-1,N}$ is a zero divisor in $R/I'$, and the sequence of the $P_{ij}$ is \emph{not} a regular sequence. Hence $R/I$ is not a complete intersection.

\subsection{An example of complete intersection}
\label{secExCI}

Let us now consider another example. Consider the ring in $3 N^2$ variables $R = \mathbb{C}[x_{ij},y_{ij},z_{ij}]$ with $i,j = 1 \dots , N$, and the ideal $I$ defined by 
\begin{equation}
\begin{array}{c}
    XY-YX=Z \, ,  \\
    X = (x_{ij})_{i,j=1 \dots , N} \, , \qquad Y = (y_{ij})_{i,j=1 \dots , N} \, , \qquad Z = (z_{ij})_{i,j=1 \dots , N} \, . 
\end{array}
\end{equation}
The ideal is generated by the polynomials $P_{ij}$ which are the matrix elements of $XY-YX-Z$. This ring is a complete intersection, because the equations $P_{ij}=0$ can be solved one by one by variable elimination, simply solving for $z_{ij}$. In other words, all the partial ideals are just polynomial rings, in which there are no divisors of zero. 

A more complex argument is needed for the ideal 
\begin{equation}
\begin{array}{c}
   XY-YX=ZZ^T  \, ,  \\
    X = (x_{ij})_{i,j=1 \dots , N} \, , \qquad Y = (y_{ij})_{i,j=1 \dots , N} \, , \qquad Z = (z_{ij})_{i,j=1 \dots , N} \, . 
\end{array}
\end{equation}
because the right-hand side $ZZ^T$ is now quadratic in the variables, and in addition it is a symmetric matrix. However one can still show that it defines a complete intersection. Instead of giving a general proof, we show in the next section how one can use a computer program to tackle this kind of problems.

\subsection{Regular Sequences with Macaulay 2}

As we saw in the previous paragraphs, one can in some cases find relation between the defining polynomials of an ideal and prove that way that the ideal is not a complete intersection, and in some other (simple) cases prove that an ideal is a complete intersection by solving systems of equations. However, the generic case is much more complicated, and involved algorithms are soon required.

There is a Macaulay2 package\footnote{It can be downloaded at the address \url{http://www2.macaulay2.com/Macaulay2/doc/Macaulay2-1.10/share/Macaulay2/Depth.m2}. } called \texttt{Depth.m2} which contains the function \texttt{regularSequenceCheck} that can be used to determine whether an ideal is a complete intersection or not. Given a list of polynomials defining the ideal, the function \texttt{regularSequenceCheck} returns the number of terms of the sequence which are regular. Therefore, the ideal is a complete intersection if and only if this number is equal to the total number of terms of the sequence. 

Let us illustrate this with the examples of the previous sections, with $N=2$. 
\begin{itemize}
    \item In the case of the first example, we use 
    \begin{verbatim}
R:=QQ[X11, X12, X21, X22, Y11, Y12, Y21, Y22];
print(regularSequenceCheck({
X21*Y12 - X12*Y21, 
X12*Y11 - X11*Y12 + X22*Y12 - X12*Y22, 
-(X21*Y11) + X11*Y21 - X22*Y21 + X21*Y22
},R));
    \end{verbatim}
    and the program returns $2$, saying as we noted in equation (\ref{zeroDiv}) that the last term is a zero divisor. 
    \item For the second example, the code becomes
    \begin{verbatim}
R:=QQ[X11, X12, X21, X22, Y11, Y12, Y21, Y22, Z11, Z12, Z21, Z22];
print(regularSequenceCheck({-(X21*Y12) + X12*Y21 - Z11^2 - Z12^2, 
 X11*Y12 - X22*Y12 + X12*(-Y11 + Y22) - Z11*Z21 - Z12*Z22, 
 -(X11*Y21) + X22*Y21 + X21*(Y11 - Y22) - Z11*Z21 - Z12*Z22, 
 X21*Y12 - X12*Y21 - Z21^2 - Z22^2
},R));
    \end{verbatim}
    and now the answer is $4$, showing that the sequence is regular and the ideal will define a complete intersection. 
\end{itemize}

\section{Proof of formula (\ref{HS1})}
\label{AppendixProof}

We want to compute the Hilbert series 
   \begin{equation}
      H_N (t,u) = \prod\limits_{i=1}^N \oint_{\mid z_i \mid =1} \frac{\d z_i}{ 2 \pi i z_i} \mathrm{PE} \left[-N t^2 + t \left(\frac{z_1}{u} + \frac{u}{z_1}\right) + t \sum\limits_{i\in \mathbb{Z}_N} \left(\frac{z_i}{z_{i+1}} + \frac{z_{i+1}}{z_{i}} \right) \right] \, .
    \end{equation}
    For that, we define the partially integrated functions
    \begin{eqnarray}
         H_{N,r} (t,u, z_1 , \dots , z_{N-r}) &=& \\ \nonumber 
         & &  \kern-10em \prod\limits_{i=N+1-r}^N \oint_{\mid z_i \mid=1}  \frac{\d z_i}{ 2 \pi i z_i} \mathrm{PE}\left[-N t^2 + t \left(\frac{z_1}{u} + \frac{u}{z_1}\right) + t \sum\limits_{i\in \mathbb{Z}_N} \left(\frac{z_i}{z_{i+1}} + \frac{z_{i+1}}{z_{i}} \right) \right] \, ,
    \end{eqnarray}
so that $H_N (t,u)= H_{N,N} (t,u)$, with $r=0,...,N$. The key step is then to prove by recursion that for $0 \leq r \leq N-1$, we have 
    \begin{eqnarray}
        H_{N,r}(t,u,z_1,...,z_{N-r}) &=& \mathrm{PE} \left[t \sum\limits_{i = 1}^{N-r-1} \left(\frac{z_i}{z_{i+1}} + \frac{z_{i+1}}{z_{i}} \right) -(N-r-1)t^2 \right. \\ \nonumber
        & & \left. \qquad -t^{2r+2} +t \left(\frac{z_1}{u} + \frac{u}{z_1}\right)  + t^{r+1} \left(\frac{z_1}{z_{N-r}} + \frac{z_{N-r}}{z_1}\right) \right] \, .
    \end{eqnarray}
Then it suffices to evaluate at $r=N-1$ to obtain an integral with only one variable which is straightforward to evaluate,  
    \begin{equation}
        H_{N}(t,u)  = \oint_{\mid z_1 \mid =1}  \frac{\d z_1}{ 2 \pi i z_1} \mathrm{PE} \left[-t^{2N}  +t \left(\frac{z_1}{u} + \frac{u}{z_1}\right)  +2 t^{N}  \right] = \mathrm{PE}[t^2+2t^N-t^{2N}] \, . 
    \end{equation}  
We note in passing that the fugacity for the global symmetry $u$ disappears during the last integration. 
    
\end{appendix}
\bibliographystyle{JHEP}
\bibliography{bibli.bib}
\end{document}